\documentclass[prd,nofootinbib]{revtex4}
\usepackage{epsfig,amsmath,amssymb,slashed}

\newcommand{\di}{{\rm d}}
\newcommand{\ii}{i}

\def\wT{{\widehat T}}
\def\wj{{\widehat j}}
\def\wspt{{\widehat{\cal S}}}

\def\wPhi{{\widehat{\Phi}}}
\def\wXi{{\widehat{\Xi}}}

\def\wrho{{\widehat{\rho}}}
\def\wUps{{\widehat{\Upsilon}}}
\def\wUpsl{{\widehat{\Upsilon}_{\rm LE}}}
\def\dwUps{\delta{\widehat{\Upsilon}}}
\def\dwUpsl{\delta{\widehat{\Upsilon}_{\rm LE}}}
\def\wvphi{{\widehat{\varphi}}}
\def\wX{{\widehat{X}}}

\def\codevc{{\stackrel{\leftrightarrow}{\nabla}}}
\newcommand{\tr}{{\rm tr}}  
  
\newcommand{\e}{{\rm e}}

\newcommand{\x}{{\rm x}}

\newcommand{\Psibar}{{\overline \Psi}}
\newcommand{\be}{\begin{equation}}
\newcommand{\ee}{\end{equation}}                                                                               
\newcommand{\bea}{\begin{eqnarray}}
\newcommand{\eea}{\end{eqnarray}}
\newcommand{\betaeq}{\beta^{\rm eq}}                                                               
\newcommand{\xieq}{\xi^{\rm eq}}  
\newcommand{\omegaeq}{\omega^{\rm eq}}                                                              

\newcommand{\brackets}[1]{\left(#1\right)}
\newcommand{\comm}[1]{\left[#1\right]}

\newcommand{\dens}{\widehat \rho}
\newcommand{\denseql}{\dens_{\rm LE}}
\newcommand{\diff}{{\rm d}}

\newcommand{\ex}[1]{{\rm e}^{#1}}
\newcommand{\betabar}{{\bar \beta}}

\newcommand{\deltabeta}{\delta\!\beta}

\newcommand{\trace}[1]{{\rm tr}\brackets{#1}}
\newcommand{\ave}[1]{\langle #1 \rangle}
\newcommand{\aveq}[1]{\ave{#1}_{0}}
\newcommand{\aveql}[1]{\ave{#1}_{\rm LE}}
\newcommand{\aveqU}[1]{\ave{#1}_{\widehat\Upsilon}}
\newcommand{\obs}{{\widehat{O}}}

\begin{document}

\title{Nonequilibrium Thermodynamical Inequivalence of Quantum Stress-energy 
and Spin Tensors} 

\author{F. Becattini, L. Tinti}\affiliation{Universit\`a di 
 Firenze and INFN Sezione di Firenze, Florence, Italy} 

\begin{abstract}
It is shown that different pairs of stress-energy and spin tensors of quantum 
relativistic fields related by a pseudo-gauge transformation, i.e. differing by
a divergence, imply different mean values of physical quantities in thermodynamical
nonequilibrium situations. Most notably, transport coefficients and the total 
entropy production rate are affected by the choice of the spin tensor of the 
relativistic quantum field theory under consideration. Therefore, at least in 
principle, it should be possible to disprove a fundamental stress-energy tensor and/or
to show that a fundamental spin tensor exists by means of a dissipative thermodynamical 
experiment.
\end{abstract}

\maketitle

\section{Introduction}
\label{intro}

In recent years, there has been a considerable interest in theoretical relativistic 
hydrodynamics and its most general form including dissipative terms \cite{various}. 
This renewed interest has been mainly triggered by its successful application to the 
description of the Quark Gluon Plasma dynamical evolution in ultreralativistic heavy 
ion collisions \cite{variousqgp}. Relativistic hydrodynamics can be seen as the theory 
describing the dynamical behaviour of the mean value of the quantum stress-energy tensor 
$\wT^{\mu\nu}$, that is $\tr (\wrho \wT^{\mu\nu})$. This tensor is generally assumed 
to be symmetric, although in {\em special} relativity it does not need to be such 
if it is accompanied by a non-vanishing rank 3 tensor, the so-called spin tensor 
$\wspt^{\lambda,\mu\nu}$. 
In fact, in special relativistic quantum field theory, starting from particular 
stress-energy and spin tensors, different pairs can be generated (and are generally 
related) by means of a pseudo-gauge transformation \cite{halb,hehl} preserving the total 
energy, momentum and angular momentum:
\begin{eqnarray}\label{transfq}
 && \wT'^{\mu \nu} = \wT^{\mu \nu} +\frac{1}{2} \partial_\alpha
 \left( \wPhi^{\alpha, \mu \nu } - \wPhi^{\mu, \alpha \nu} - 
 \wPhi^{\nu, \alpha \mu}  \right) \nonumber \\
 && \wspt'^{\lambda, \mu \nu} = \wspt^{\lambda,\mu\nu}-\wPhi^{\lambda,\mu\nu}
 + \partial_\alpha \widehat Z^{\alpha \lambda,\mu \nu}
\end{eqnarray}
where $\wPhi$ is a rank three tensor field antisymmetric in the last two indices
(often called and henceforth referred to as {\em superpotential}) and $\widehat Z$ 
a rank four tensor antisymmetric in the pairs $\alpha \lambda$ and $\mu \nu$.

In a previous paper \cite{bt1} we have shown that indeed different pairs $(\wT,\wspt)$ 
and $(\wT',\wspt')$ are in general thermodynamically inequivalent as they imply different 
mean values of physical quantities for a rotating system at equilibrium. Particularly, 
for the free Dirac field, we showed that the canonical and Belinfante (obtained from 
the canonical one by setting $\wPhi= \wspt$ and $\widehat Z = 0$ in (\ref{transfq}), hence with 
a vanishing new spin tensor $\wspt'$) quantum stress-energy tensors result in different 
mean values for the momentum density and the total angular momentum density. 

The thermodynamical inequivalence is (at least in our view) surprising because it 
was commonly believed that the only physical phenomenon which can discriminate between 
stress-energy tensors of a fundamental quantum field theory related by a transformation 
like (\ref{transfq}) is gravity, or, in other words, the coupling to a metric tensor. 
In this paper we reinforce our previous finding by showing that the inequivalence 
extends to nonequilibrium thermodynamical quantities, 
specifically entropy production and transport coefficients. In summary, we will show 
that the use of different stress-energy tensors, related by (\ref{transfq}), to 
calculate transport coefficients with the relativistic Kubo formula leads, in 
general, to different results. Therefore, at least in principle, an extremely accurate 
measurement of transport coefficients or total entropy in an experiment where 
dissipation is involved, would allow to {\em disprove} a candidate stress-energy or 
spin tensor, with obvious important consequences in relativistic gravitational theories. 
This finding means, in other words, that the existence of a fundamental spin tensor 
affects the microscopic number of degrees of freedom, or at least on how quickly
macroscopic information gets converted into microscopic, namely on entropy generation.

The paper is organized as follows: in Sect.~\ref{zuba} we will extend the framework
of the nonequilibrium density operator introduced by Zubarev \cite{zubarev} to
the case of a non-vanishing spin tensor. In Sect.~\ref{ineq}, it will be shown that
the nonequilibrium density operator is not invariant under a pseudo-gauge transformation
(\ref{transfq}), that is it does depend on the chosen couple of stress-energy 
and spin tensor. In Sect.~\ref{mean} we will provide a general formula for the
change of mean values of observables and we will determine how entropy is affected
by a pseudo-gauge transformation. In Sect.~\ref{kubo} we will show that transport
coefficients are also modified and, particularly, we will focus on the modification
of the Kubo formula for shear viscosity. Finally, in Sect.~\ref{discu}, we will
discuss the implications of this finding and draw our conclusions.

\subsection*{Notation}

In this paper we adopt the natural units, with $\hbar=c=K=1$.\\
The Minkowskian metric tensor is ${\rm diag}(1,-1,-1,-1)$; for the Levi-Civita
symbol we use the convention $\varepsilon^{0123}=1$.\\ 
We will use the relativistic notation with repeated indices assumed to 
be saturated. Operators in Hilbert space will be denoted by an upper hat, e.g. 
$\widehat {\sf R}$, with the exception of the Dirac field operator which is 
denoted with a capital $\Psi$.

\section{Nonequilibrium density operator}
\label{zuba}

A suitable formalism to calculate transport coefficients for relativistic quantum
fields without going through kinetic theory was developed by Zubarev \cite{zubarev,zubabooks},
extending to the relativistic domain a formalism already introduced by Kubo \cite{kubolibro}. 
In this approach, a non-equilibrium density operator is introduced which reads \cite{hosoya}
\footnote{Throughout the paper, the four-vector $x$ implies the time $t$ and position
vector ${\bf x}$, i.e. $x=(t,{\bf x})$. The dependence of the stress-energy and spin 
tensor on $x$ will always be understood.}:
\be\label{zub1}
 \wrho = \frac{1}{Z} \exp [ - \wUps] = 
 \frac{1}{Z}  \exp \left[ -\lim_{\varepsilon \to 0} \varepsilon \int^{t'}_{-\infty} 
 \!\!\! \di t \; \e^{\varepsilon (t-t')} \int \di^3 \x \; \left( \wT^{0\nu} \beta_\nu(x)
 - \wj^0 \xi(x) \right) \right]
\ee
where $\wj$ is a conserved current, the four-vector field $\beta$ is a point-dependent 
inverse temperature four-vector ($\beta = u/T_0$, $u$ being a four-velocity field 
and $T_0$ the comoving or invariant temperature) and $\xi = \mu_0/T_0$ a scalar 
function whose physical meaning 
is that of a point-dependent ratio between comoving chemical potential $\mu_0$ and 
comoving temperature $T_0$; the $Z$ factor is analogous to a partition function, i.e.
a normalization factor to have $\tr \wrho = 1$. The operators in the exponential of 
Eq.~(\ref{zub1}) are in the Heisenberg representation. It should be stressed that
in the formula (\ref{zub1}) covariance is broken from the very beginning by the 
choice of a specific inertial frame and its time. However, it can be shown 
that the operator $\wrho$ is in fact time-independent \cite{hosoya}, namely independent 
of $t'$, so that $\wrho$ is a good density operator in the Heisenberg representation. 

In the formula (\ref{zub1}) the possible contribution of a spin tensor is simply
disregarded; therefore, the formula is correct only if the stress-energy tensor 
is the symmetrized Belinfante one (or improved ones, see last section), whose 
associated spin tensor is vanishing. It is the aim of this Section to find the 
appropriate extension of the formula (\ref{zub1}) with a spin tensor.

Using the identity:
$$
  \e^{\varepsilon (t-t')} \left( \wT^{0\nu} \beta_\nu(x)
  - \wj^0 \xi(x) \right) = \left( \frac{\partial}{\partial x^\mu}
  \frac{\e^{\varepsilon (t-t')}}{\varepsilon} \right) 
  \left( \wT^{\mu\nu} \beta_\nu(x) - \wj^\mu \xi(x) \right)    
$$
integrating by parts and taking into account the continuity equations $\partial_\mu 
\wT^{\mu \nu} = \partial_\mu \wj^\mu = 0$, the operator $\wUps$ in Eq.~(\ref{zub1}) 
can be rewritten as:
\bea\label{elabor}
 \wUps &=& \int \di^3 \x \; \left( \wT^{0\nu} \beta_\nu(t',{\bf x})
 - \wj^0 \xi(t',{\bf x}) \right) + \lim_{\varepsilon \to 0} 
 \int^{t'}_{-\infty} \!\!\! \di t \; \e^{\varepsilon (t-t')} \int  \di S \, n_i 
 \left( \wT^{i \nu} \beta_\nu(x) - \wj^i \xi(x) \right) \nonumber \\
 &-& \lim_{\varepsilon \to 0} 
\int^{t'}_{-\infty} \!\!\! \di t \; \e^{\varepsilon (t-t')} \int \di^3 \x \; 
\left( \wT^{\mu \nu} \partial_\mu \beta_\nu(x) - \wj^\mu \partial_\mu \xi(x) \right)
\eea
The first term the so-called {\em local thermodynamical equilibrium} one, which is defined 
by the same formula of the global equilibrium \cite{weldon,becatens} with $x$-dependent 
four-temperature and chemical potentials, whereas the term dependent on their 
derivatives is interpreted as a perturbation. 

At equilibrium, the right hand side should reduce to the known form, which, at least 
for the most familiar form of thermodynamical equilibrium with $\betaeq = (1/T,{\bf 0}) 
= const$ and $\xieq = \mu/T = const$ is readily recognized in the first term setting
$\beta=\betaeq$ and $\xi=\xieq$:
\bea\label{upseq}
 \wUps^{\rm eq} &=& 
 \int \di^3 \x \; \left( \wT^{0\nu} \betaeq_\nu - \wj^0 \xieq \right) + 
 \lim_{\varepsilon \to 0} 
 \int^{t'}_{-\infty} \!\!\! \di t \; \e^{\varepsilon (t-t')} \int \di S \, n_i 
 \left( \wT^{i \nu} \betaeq_\nu - \wj^i \xieq \right) \nonumber \\
 &-& \lim_{\varepsilon \to 0} 
\int^{t'}_{-\infty} \!\!\! \di t \; \e^{\varepsilon (t-t')} \int \di^3 \x \;
\left( \wT^{\mu \nu} \partial_\mu \betaeq_\nu - \wj^\mu \partial_\mu \xieq \right)
  = \widehat H/T - \mu \widehat Q/T \nonumber \\
  &+& \lim_{\varepsilon \to 0} 
 \int^{t'}_{-\infty} \!\!\! \di t \; \e^{\varepsilon (t-t')} \int \di S \, n_i 
 \left( \wT^{i \nu} \betaeq_\nu - \wj^i \xieq \right) - \lim_{\varepsilon \to 0} 
  \int^{t'}_{-\infty} \!\!\! \di t \; \e^{\varepsilon (t-t')} \int \di^3 \x \;
 \left( \wT^{\mu \nu} \partial_\mu \betaeq_\nu - \wj^\mu \partial_\mu \xieq \right)
\eea
Hence, the two rightmost terms of (\ref{upseq}) must vanish at equilibrium. Indeed, 
the surface term is supposed to vanish through a suitable choice of the field boundary 
conditions while the third term vanishes in view of the constancy of $\betaeq$ and 
$\xieq$. However, this is not the case for the most general form of equilibrium;
in the most general form (see discussion in ref.~\cite{becatens}), whilst the scalar 
$\xieq$ stays constant the four-vector $\beta$ fulfills a Killing equation, whose 
solution is \cite{degroot}:
\be\label{geneq}
  \betaeq_\nu(x) = b_\nu^{\rm eq} + \omegaeq_{\nu \mu} x^\mu
\ee
with both the four-vector $b^{\rm eq}$ and the antisymmetric tensor $\omega^{\rm eq}$ 
constant. Therefore:
$$
  \partial_\mu \beta_\nu^{\rm eq} = - \omegaeq_{\mu \nu}  
$$
which in general is non-vanishing, so that the third term on the right hand side of
Eq.~(\ref{upseq}) survives. For instance, for the thermodynamical equilibrium with 
rotation \cite{becatens}, the tensor $\omega$ turns out to be:
\be\label{omegat}
  \omega^{\rm eq}_{\lambda \nu} = \omega/T \left( \delta^1_\lambda \delta^2_\nu
  -\delta^2_\lambda \delta^1_\nu \right)
\ee
$\omega$ being the angular velocity and $T$ the temperature measured by the 
inertial frame. 

In order to find the appropriate generalization of the operator $\wUps$, let us 
plug the formula (\ref{geneq}) of general thermodynamical equilibrium into the 
(\ref{upseq}):
\bea\label{upseqgen}
 \wUps^{\rm eq} &=& 
 \int \di^3 \x \; \left( \wT^{0\nu} \betaeq_\nu - \wj^0 \xieq) \right) + 
 \lim_{\varepsilon \to 0} \int^{t'}_{-\infty} \!\!\! \di t \; \e^{\varepsilon (t-t')}
 \int \di S \, n_i \left( \wT^{i \nu} (b^{\rm eq}_\nu + \omegaeq_{\nu \mu} 
 x^\mu) - \wj^i \xieq \right) \nonumber \\
 &+& \lim_{\varepsilon \to 0} \int^{t'}_{-\infty} \!\!\! \di t \; \e^{\varepsilon (t-t')}
  \int \di^3 \x \; \wT^{\mu \nu} \omegaeq_{\mu \nu}
\eea
where $\partial_\mu \xieq = 0$ has been taken into account. For a symmetric 
stress-energy tensor $\wT$, the last term vanishes, but if a spin tensor is present $\wT$
may have an antisymmetric part. Particularly, from the angular momentum continuity
equation:
\be
  \wT^{\mu \nu} \omegaeq_{\mu \nu} = \frac{1}{2} (\wT^{\mu \nu}-\wT^{\nu \mu})
  \omegaeq_{\mu \nu} = - \frac{1}{2} \partial_\lambda \wspt^{\lambda,\mu\nu}
  \omegaeq_{\mu \nu}
\ee
so that the last term on the right hand side of Eq.~(\ref{upseqgen}) can be rewritten 
as:
\bea\label{thirdt}
 && \lim_{\varepsilon \to 0} \int^{t'}_{-\infty} \!\!\! \di t \; \e^{\varepsilon (t-t')}
 \int \di^3 \x \;  \wT^{\mu \nu} \omegaeq_{\mu \nu} = 
 - \frac{1}{2} \omegaeq_{\mu \nu} \lim_{\varepsilon \to 0} 
 \int^{t'}_{-\infty} \!\!\! \di t \; \e^{\varepsilon (t-t')} \int \di^3 \x \; 
 \partial_\lambda \wspt^{\lambda,\mu\nu} \nonumber \\
 && = - \frac{1}{2} \omegaeq_{\mu \nu} \lim_{\varepsilon \to 0} \int \di^3 \x 
  \int^{t'}_{-\infty} \!\!\! \di t \; \e^{\varepsilon (t-t')} \frac{\partial}{\partial t} 
  \wspt^{0,\mu\nu} - \frac{1}{2} \omegaeq_{\mu \nu} \lim_{\varepsilon \to 0} 
  \int^{t'}_{-\infty} \!\!\!  \di t \; \e^{\varepsilon (t-t')} \int \di S \, n_i  
  \wspt^{i,\mu\nu}
\eea
The first term on the right hand side of (\ref{thirdt}) can be integrated by parts, yielding:
\be\label{fourtht}
- \frac{1}{2} \omegaeq_{\mu \nu} \lim_{\varepsilon \to 0} \int \di^3 \x 
  \int^{t'}_{-\infty} \!\!\! \di t \; \e^{\varepsilon (t-t')} \frac{\partial}{\partial t} 
  \wspt^{0,\mu\nu} = - \frac{1}{2} \omegaeq_{\mu \nu} \int \di^3 \x \, 
  \wspt^{0,\mu\nu} (t',{\bf x}) + \frac{1}{2} \omegaeq_{\mu \nu} 
  \lim_{\varepsilon \to 0} \varepsilon \int^{t'}_{-\infty} \!\!\! \di t 
  \; \e^{\varepsilon (t-t')} \int \di^3 \x \; \wspt^{0,\mu\nu} (x)
\ee  
Plugging the Eq.~(\ref{fourtht}) into (\ref{thirdt}) and this in turn into (\ref{upseqgen})
we obtain:
\bea\label{upseqgen2}
 \wUps^{\rm eq} &=& 
 \int \di^3 \x \; \left( \wT^{0\nu} \betaeq_\nu - \wj^0 \xieq - \frac{1}{2} \omegaeq_{\mu \nu}
 \wspt^{0, \mu \nu} \right) + \lim_{\varepsilon \to 0} \int^{t'}_{-\infty} \!\!\! \di t \; 
 \e^{\varepsilon (t-t')} \left[ b^{\rm eq}_\nu \int \di S \, n_i \wT^{i \nu} - \xieq 
 \int \di S \, n_i \wj^{i} \right. \nonumber \\
 &-& \left. \frac{1}{2} \omegaeq_{\mu \nu} \int \di S \, n_i ( x^\mu \wT^{i \nu} - 
 x^\nu \wT^{\mu i} + \wspt^{i,\mu \nu} ) \right] + \frac{1}{2} \omegaeq_{\mu \nu} 
  \lim_{\varepsilon \to 0} \varepsilon \int^{t'}_{-\infty} \!\!\! \di t \; 
   \e^{\varepsilon (t-t')} \int \di^3 \x \; \wspt^{0,\mu\nu} (x)
\eea
where the surface term involving $\wT$ in Eq.~(\ref{upseqgen}) has been rearranged
taking advantage of the antisymmetry of the $\omega$ tensor. The surface terms in
the above equations now are manifestly the total momentum flux, the charge flux and
the {\em total} angular momentum flux through the boundary. All of these terms are
supposed to vanish at thermodynamical equilibrium through suitable conditions 
enforced on the field operators at the boundary, so that the (\ref{upseqgen2}) reduces 
to:
\be\label{upseqgen3}
 \wUps^{\rm eq} = 
 \int \di^3 \x \; \left( \wT^{0\nu} \betaeq_\nu - \wj^0 \xieq - \frac{1}{2} \omegaeq_{\mu \nu}
 \wspt^{0, \mu \nu} \right) + \frac{1}{2} \omegaeq_{\mu \nu} 
  \lim_{\varepsilon \to 0} \varepsilon \int^{t'}_{-\infty} \!\!\! \di t
   \; \e^{\varepsilon (t-t')}\int \di^3 \x  \; \wspt^{0,\mu\nu} (x)
\ee
The first term on the right hand side just gives rise to the desired form of the 
equilibrium operator. For instance, for a rotating system with $\omega$ as in 
Eq.~(\ref{omegat}) one has \cite{becatens}:
$$
 \int \di^3 \x \; \left( \wT^{0\nu} \betaeq_\nu - \wj^0 \xieq - \frac{1}{2} \omegaeq_{\mu \nu}
 \wspt^{0, \mu \nu} \right) = {\widehat H}/T - \mu {\widehat Q}/T - 
 \omega {\widehat J}/T
$$
$\widehat J$ being the total angular momentum, which is the known form \cite{balian}.
Nevertheless, the second term in Eq.~(\ref{upseqgen3}) does not vanish and, thus,
must be subtracted away with a suitable modification of the definition of the 
$\wUps$ operator. The form of the unwanted term demands the following modification 
of (\ref{zub1}):
\be\label{zub2}
 \wrho = \frac{1}{Z} \exp [ - \wUps] = 
 \frac{1}{Z} \exp \left[ -\lim_{\varepsilon \to 0} \varepsilon \int^{t'}_{-\infty} 
 \!\!\! \di t \; \e^{\varepsilon (t-t')}\int \di^3 \x \;  \left( \wT^{0\nu} \beta_\nu(x)
 - \wj^0 \xi(x) - \frac{1}{2} \wspt^{0,\mu\nu} \omega_{\mu \nu}(x) \right) \right]
\ee
where $\omega_{\mu \nu}(x)$ is an antisymmetric tensor field which must reduce
to the constant $\omegaeq_{\mu \nu}$ tensor at equilibrium. It is easy to check, by 
tracing the previous calculations, that the equilibrium form of $\wUps$ reduces to 
the desired form:
$$
 \wUps^{\rm eq} = \int \di^3 \x \; \left( \wT^{0\nu} \betaeq_\nu - \wj^0 \xieq - 
 \frac{1}{2} \omegaeq_{\mu \nu} \wspt^{0, \mu \nu} \right) 
$$
as the spin tensor term in Eq.~(\ref{upseqgen3}) cancels out. Therefore, the operator 
(\ref{zub2}) is the only possible extension of the nonequilibrium density operator 
with a spin tensor. 

The new operator $\wUps$ can be worked out the same way as we have done when obtaining 
Eq.~(\ref{elabor}) from Eq.~(\ref{zub1}): 
\bea\label{elabor2}
 \wUps &=& \int \di^3 \x \; \left( \wT^{0\nu} \beta_\nu(t',{\bf x})
 - \wj^0 \xi(t',{\bf x}) -\frac{1}{2} \wspt^{0,\mu\nu} \omega_{\mu\nu}(t',{\bf x}) \right) 
 \nonumber \\
 &+& \lim_{\varepsilon \to 0} \int^{t'}_{-\infty} \!\!\! \di t \; \e^{\varepsilon (t-t')} 
 \int  \di S \, n_i \left( \wT^{i \nu} \beta_\nu(x) - \wj^i \xi(x) - \frac{1}{2} \wspt^{i,\mu\nu} 
 \omega_{\mu\nu}(x)\right) \nonumber \\
 &-& \frac{1}{2} \lim_{\varepsilon \to 0} \int^{t'}_{-\infty} \!\!\! \di t \; 
 \e^{\varepsilon (t-t')}\int \di^3 \x \; \left( \wT_S^{\mu \nu} (\partial_\mu 
 \beta_\nu(x)+\partial_\mu \beta_\nu(x))  
 + \wT_A^{\mu \nu} (\partial_\mu \beta_\nu(x) - \partial_\mu \beta_\nu(x) 
 + 2 \omega_{\mu\nu}(x)) \right. \nonumber \\
 && \left. - \wspt^{\lambda,\mu\nu} \partial_\lambda \omega_{\mu\nu}(x) - 
 2 \wj^\mu \partial_\mu \xi(x) \right) 
\eea
where:
$$
  \wT_S^{\mu \nu} = \frac{1}{2} (\wT^{\mu \nu} + \wT^{\nu \mu}) \qquad \qquad
  \wT_A^{\mu \nu} = \frac{1}{2} (\wT^{\mu \nu} - \wT^{\nu \mu})
$$  
and the continuity equation for angular momentum has been used.
The first term on the right hand side is the new local thermodynamical term whilst 
the third term can be further expanded to derive the relativistic Kubo formula
of transport coefficients (see Appendix A).

\section{Nonequilibrium density operator and pseudo-gauge transformations}
\label{ineq}

A natural requirement for the density operator (\ref{zub2}) would be its independence of
the particular couple of stress-energy and spin tensor, because one would like the 
mean value of any observable $\widehat O$ :
$$
 O \equiv \tr (\wrho \, \obs)
$$
to be an objective one \footnote{It should be pointed out that the mean value 
of operators involving quantum relativistic fields are generally divergent (e.g. 
$T^{00}$ for a free field has an infinite zero point value). To remove the infinities, 
the mean values must be renormalized, what can be simply done for free fields by using
normal ordering in all expressions, including the density operator itself. Henceforth,
it will be understood that all the mean values of operators are the renormalized ones.}
. In ref.~\cite{bt1} we showed that even at thermodynamical 
equilibrium with rotation this is not the case for the components of the stress-energy 
and spin tensor themselves because they change through the pseudo-gauge transformation 
(\ref{transfq}). However, at equilibrium, $\wrho$ itself is a function of just integral 
quantities (total energy, angular momentum, charge) which are invariant under a 
transformation (\ref{transfq}) provided that boundary fluxes vanish, so a specific 
operator $\obs$, including the components of a {\em specific} stress-energy tensor, 
does not change under (\ref{transfq}). However, it is not obvious that this feature 
persists in a nonequilibrium case, in fact we are going to show that, in general, 
this is not the case.

Let us consider the operator $\wUps$ in (\ref{zub2}) and how it gets changed under
a pseudo-gauge transformation (\ref{transfq}) with $\widehat Z=0$. The new operator
$\wUps'$ reads:
\be\label{deltau1}
 \wUps'= \wUps + \frac{1}{2} \lim_{\varepsilon \to 0} \varepsilon \int^{t'}_{-\infty} 
 \!\!\! \di t \; \e^{\varepsilon (t-t')} \int \di^3 \x \; \left( \partial_\lambda 
 \wvphi^{\lambda 0,\nu} \beta_\nu(x) + \wPhi^{0,\mu \nu} \omega_{\mu \nu}(x) \right)
\ee
where: 
\be\label{phidef}
 \wvphi^{\lambda \mu, \nu} = \wPhi^{\lambda, \mu \nu } - \wPhi^{\mu, \lambda \nu} - 
 \wPhi^{\nu, \lambda \mu} 
\ee
is antisymmetric in the first two indices. We can rewrite Eq.~(\ref{deltau1}) as:
\bea\label{deltau2}
 \wUps' -\wUps &=& \frac{1}{2} \lim_{\varepsilon \to 0} \varepsilon \int^{t'}_{-\infty} 
 \!\!\! \di t \int \di^3 \x \; \e^{\varepsilon (t-t')} \left[ \partial_\lambda 
 (\wvphi^{\lambda 0,\nu} \beta_\nu(x)) - \wvphi^{\lambda 0,\nu} \partial_\lambda 
 \beta_\nu + \wPhi^{0,\mu \nu} \omega_{\mu \nu}(x) \right] \nonumber \\
 &=& \frac{1}{2} \lim_{\varepsilon \to 0} \varepsilon \int^{t'}_{-\infty} 
 \!\!\! \di t \; \e^{\varepsilon (t-t')} \left[ \int \di S \, n_i \, \wvphi^{i 0,\nu} 
 \beta_\nu(x) - \int \di^3 \x \; \left( \wvphi^{\lambda 0,\nu} \partial_\lambda 
 \beta_\nu - \wPhi^{0,\mu \nu} \omega_{\mu \nu}(x) \right) \right]
\eea
after integration by parts. Let us now write the general fields $\beta$ and $\omega$ 
as the sum of the equilibrium values and a perturbation, that is:
\be
   \beta(x) = \betaeq(x) + \delta \beta(x)  \qquad \qquad \qquad 
   \omega(x) = \omegaeq + \delta \omega(x)
\ee
and work out first the equilibrium part of the right hand side of Eq.~(\ref{deltau2}).
As $\partial_\lambda \betaeq_\nu = -\omegaeq_{\lambda \nu}$ one has:
\bea\label{deltaueq}
 (\wUps' -\wUps)|_{\rm eq} &=& \frac{1}{2} 
 \lim_{\varepsilon \to 0} \varepsilon \int^{t'}_{-\infty} 
 \!\!\! \di t \; \e^{\varepsilon (t-t')} \left[ \int \di S \, n_i \, \wvphi^{i 0,\nu} 
 \betaeq_\nu(x) + \int \di^3 \x \; \left( \wvphi^{\lambda 0,\nu} \omegaeq_{\lambda \nu}
 + \wPhi^{0,\mu \nu} \omegaeq_{\mu \nu} \right) \right] \nonumber \\
 &=& \frac{1}{2} \lim_{\varepsilon \to 0} \varepsilon \int^{t'}_{-\infty} 
 \!\!\! \di t \; \e^{\varepsilon (t-t')} \left[ \int \di S \, n_i \, \wvphi^{i 0,\nu} 
 \betaeq_\nu(x) + \int \di^3 \x \; \left( \wPhi^{\lambda, 0 \nu } \omegaeq_{\lambda \nu} - 
  \wPhi^{0, \lambda \nu} \omegaeq_{\lambda \nu} - \wPhi^{\nu, \lambda 0} 
  \omegaeq_{\lambda \nu} + \wPhi^{0,\mu \nu} \omegaeq_{\mu \nu} \right) \right] \nonumber \\
 &=& \frac{1}{2} \lim_{\varepsilon \to 0} \varepsilon \int^{t'}_{-\infty} 
 \!\!\! \di t \; \e^{\varepsilon (t-t')} \int \di S \, n_i \wvphi^{i 0,\nu} 
  \betaeq_\nu(x)
\eea
where we have used the Eq.~(\ref{phidef}) and the antisymmetry of indices of the 
superpotential $\wPhi$. By using the Eq.~(\ref{geneq}), the last expression can 
be rewritten as:
$$
\lim_{\varepsilon \to 0} \varepsilon \int^{t'}_{-\infty} 
 \!\!\! \di t \; \e^{\varepsilon (t-t')} \left[ b^{\rm eq}_\nu \int \di S \, n_i 
  \wvphi^{i 0,\nu} + \frac{1}{2} \omegaeq_{\nu \mu} \int \di S \, n_i 
  (x^\mu \wvphi^{i 0,\nu} - x^\nu \wvphi^{i 0,\mu}) \right]
$$
The two surface integrals above are the additional four-momentum and the additional
{\em total} angular momentum, in the operator sense, after having made a pseudo-gauge
tranformation (\ref{transfq}) of the stress-energy and spin tensor. If the boundary 
conditions ensure that the momentum and total angular momentum fluxes vanish (in order 
to have conserved energy and momentum operators) for any couple $(\wT,\wspt)$ of tensors, 
then the two fluxes in the above equations must vanish as well. Therefore, we can 
conclude that: 
$$
 \wUps'|_{\rm eq} =  \wUps |_{\rm eq}
$$
Now, let us focus on the nonequilibrium perturbation of the $\wUps$ operator.
\bea\label{deltanoneq}
 && (\wUps' -\wUps )|_{\rm non-eq} = \frac{1}{2} \lim_{\varepsilon \to 0} \varepsilon 
 \int^{t'}_{-\infty} \!\!\! \di t \; \e^{\varepsilon (t-t')} \left[ \int \di S 
 \, n_i \, \wvphi^{i 0,\nu} \delta \beta_\nu - \int \di^3 \x \; \wvphi^{\lambda 0,\nu} 
 \partial_\lambda \delta \beta_\nu - \wPhi^{0,\mu \nu} \delta \omega_{\mu \nu} \right]
 \nonumber \\
 && =  \frac{1}{2} \lim_{\varepsilon \to 0} \varepsilon 
 \int^{t'}_{-\infty} \!\!\! \di t \; \e^{\varepsilon (t-t')} \left[ \int \di S 
 \, n_i \, \wvphi^{i 0,\nu} \delta \beta_\nu - \int \di^3 \x \; (\wPhi^{\lambda, 0 \nu} 
 - \wPhi^{0, \lambda \nu} - \wPhi^{\nu ,\lambda 0}) \partial_\lambda \delta \beta_\nu  
 - \wPhi^{0,\mu \nu} \delta \omega_{\mu \nu} \right] \nonumber \\
 && =  \frac{1}{2} \lim_{\varepsilon \to 0} \varepsilon 
 \int^{t'}_{-\infty} \!\!\! \di t \; \e^{\varepsilon (t-t')} \left[ \int \di S 
 \, n_i \, \wvphi^{i 0,\nu} \delta \beta_\nu - \int \di^3 \x \; \wPhi^{\lambda, 0 \nu}
 (\partial_\lambda \delta \beta_\nu + \partial_\nu \delta \beta_\lambda) 
 - \wPhi^{0, \lambda \nu} \left( \frac{1}{2} (\partial_\lambda \delta \beta_\nu - 
 \partial_\nu \delta \beta_\lambda)  + \delta \omega_{\lambda \nu} \right) \right]
 \nonumber \\ 
\eea
where the dependence of $\delta\beta$ and $\delta\omega$ on $x$ is now understood.
It can be seen that it is impossible to make this difference vanishing in general.
One can get rid of the surface term by choosing a perturbation which vanishes at
the boundary and the last term by locking the perturbation of the tensor $\omega$ 
to that of the inverse temperature four-vector:
\be\label{ombetalock}
 \delta \omega_{\lambda \nu}(x) = - \frac{1}{2} (\partial_\lambda \delta \beta_\nu(x) - 
 \partial_\nu \delta \beta_\lambda(x))
\ee
but it is impossible to cancel out the term:
\be\label{deltaups}
 \dwUps \equiv  -\frac{1}{2} \lim_{\varepsilon \to 0} \varepsilon 
 \int^{t'}_{-\infty} \!\!\! \di t \; \e^{\varepsilon (t-t')} 
 \int \di^3 \x \; \wPhi^{\lambda, 0 \nu}
 (\partial_\lambda \delta \beta_\nu(x) + \partial_\nu \delta \beta_\lambda(x)) 
\ee
unless in special cases, e.g. when the tensor $\wPhi$ is also antisymmetric in the 
first two indices.

We have thus come to the conclusion that the nonequilibrium density operator does 
depend, in general, on the particular choice of stress-energy and spin tensor of the 
quantum field theory under consideration. Therefeore, the mean value of any observable 
in a non-equilibrium situation shall depend on that choice. It is worth stressing 
that this is a much deeper dependence on the stress-energy and spin 
tensor than what we showed in ref.~\cite{bt1} for thermodynamical equilibrium with 
rotation. Therein, mean values of the angular momentum densities and momentum densities 
were found to be dependent on the pseudo-gauge transformation (\ref{transfq}) because 
the relevant quantum operators could be varied, but not because the density operator 
$\wrho$ was dependent thereupon. In fact, at non-equilibrium, even $\wrho$ varies 
under a transformation (\ref{transfq}).
Note that, in principle, even the mean values of the total energy and momentum could
be dependent on the quantum stress-energy tensor choice although boundary conditions ensure, 
as we have assumed, that the total energy and momentum {\em operators} are invariant 
under a transformation (\ref{transfq}). Again, this comes about because the density 
operator is not invariant under (\ref{transfq}), in formula:
$$
 \tr (\wrho' \widehat P'^\mu) =  \tr (\wrho' \widehat P^\mu) \ne \tr (\wrho \widehat P^\mu) 
$$

It must be pointed out that the variation of the Zubarev non-equilibrium density
operator (\ref{deltaups}) depends on the gradients of the four-temperature field
and it is thus a small one close to thermodynamical equilibrium.
In the next Section we will show in more details how the mean values of observables 
change under a small change of the nonequilibrium density operator, or, in other
words, when the system is close to thermodynamical equilibrium.

\section{Variation of mean values and linear response}
\label{mean}

We will first study the general dependence of the mean value of an observable $\widehat O$ 
on the spin tensor by denoting by $\dwUps$ the supposedly small variation, under a 
transformation (\ref{transfq}), of the operator $\wUps$. This can be either the one 
in Eq.~(\ref{deltaups}) or the more general (only bulk terms) in Eq.~(\ref{deltanoneq}). 
We have:
\be
 \tr (\wrho' \widehat O) = \frac{1}{Z'} \tr (\exp[-\wUps'] \widehat O)
 = \frac{1}{Z'} \tr (\exp[-\wUps - \dwUps] \widehat O)
\ee
being $Z' = \tr(\exp[-\wUps - \dwUps])$. We can expand in $\dwUps$ at the first 
order (Zassenhaus formula):
\bea\label{zass}
 && Z' \simeq Z - \tr (\exp[-\wUps] \dwUps) \nonumber \\
 && \tr (\exp[-\wUps - \dwUps] \widehat O) \simeq \tr \left( \exp[-\wUps](I - 
 \dwUps + \frac{1}{2} [\wUps,\dwUps] - \frac{1}{6} [\wUps,[\wUps,\dwUps]] 
 + \ldots) \widehat O \right)
\eea
hence, with $\langle \,\, \rangle = \tr (\wrho \,\,)$, at the first order in
$\dwUps$:
$$
 \tr (\wrho' \widehat O) \equiv \langle \widehat O \rangle' \simeq 
  \langle \widehat O \rangle (1 + \langle \dwUps \rangle)
 - \langle \widehat O \dwUps \rangle + \frac{1}{2} \langle [\wUps,\dwUps] 
  \widehat O \rangle - \frac{1}{6} \langle [\wUps,[\wUps,\dwUps]] \widehat O \rangle 
  + \ldots 
$$
which makes manifest the dependence of the mean value on the choice of the superpotential 
$\wPhi$. 

As has been mentioned, close to thermodynamical equilibrium, the operator $\dwUps$ 
is ``small" and one can write an expansion of the mean value of the observable $\obs$ 
in the gradients of the four-temperature field, according to relativistic linear 
response theory \cite{hosoya}. This method, just based on Zubarev's nonequilibrium 
density operator method, 
allows to calculate the variation between the actual mean value of an operator and 
its value at local thermodynamical equilibrium for small deviations from it. In fact,
it can be seen from Eq.~(\ref{deltaups}) that the operator $\dwUps$, from the linear 
response theory viewpoint, is an additional perturbation in the derivative of the 
four-temperature field and therefore the difference between actual mean values at 
first order turns out be (see Appendix A for reference):
\be\label{lrto}
 \Delta \langle \widehat O \rangle \simeq -
 \lim_{\varepsilon \to 0} \frac{T}{2 \ii} \int_{-\infty}^{t'} \di t \;
 \e^{\varepsilon(t-t')} \int \di^3 \x \; \langle [\wPhi^{\lambda,0\nu}(x),\widehat O] 
 \rangle_0 (\partial_\lambda \delta \beta_\nu (x) + \partial_\nu \delta \beta_\lambda (x)) 
\ee
where $\langle \ldots \rangle_0$ stands for the expectation value calculated with the 
equilibrium density operator, that is:
\be\label{rhoeq}
  \wrho_0 = \frac{1}{Z_0} \exp[-\widehat H/T + \mu \widehat Q/T]
\ee
Since $\tr (\wrho_0 [\wPhi^{\lambda, 0 \nu},\widehat O]) = \tr(\wPhi^{\lambda, 0 \nu}
[\widehat O,\wrho_0])$ the right hand side of (\ref{lrto}) vanishes for all quantities 
commutating with the equilibrium density operator, notably total energy, momentum and 
angular momentum. Nevertheless, in principle, even the mean 
values of the conserved quantities are affected by the choice of a specific quantum 
stress-energy tensor, though at the second order in the perturbation $\delta \beta$. 

We now set out to study the effect of the transformation (\ref{transfq}) on the total 
entropy. In nonequilibrium situation, entropy is usually defined as \cite{balian} the
quantity maximizing $-\tr (\wrho \log \wrho)$ with the constraints of fixed mean 
conserved densities. The solution $\wrho_{\rm LE}$ of this problem is the local 
thermodynamical equilibrium operator, namely:
\be\label{rholeq}
  \wrho_{\rm LE}(t) = \frac{\exp[- \int \di^3 \x \; \left( \wT^{0\nu} 
  \beta_\nu(x) - \wj^0 \xi(x) -\frac{1}{2} \wspt^{0,\mu\nu} \omega_{\mu\nu}(x) 
  \right)]}{\tr(\exp[- \int \di^3 \x \; \left( \wT^{0\nu} 
  \beta_\nu(x) - \wj^0 \xi(x) -\frac{1}{2} \wspt^{0,\mu\nu} \omega_{\mu\nu}(x) 
  \right)])}
\ee
which - as emphasized in the above equation - is explicitely dependent on time, 
unlike the Zubarev stationary nonequilibrium density operator (\ref{zub2}); of 
course the time dependence is crucial to make entropy 
\be\label{ent}
 S = -\tr (\wrho_{\rm LE} \log \wrho_{\rm LE})
\ee
increasing in nonequilibrium situation. 
In order to study the effect of the transformation (\ref{transfq}) on the entropy
it is convenient to define:
\be\label{loceq}
 \wUpsl = \int \di^3 \x \; \left( \wT^{0\nu} \beta_\nu(x) - \wj^0 \xi(x) 
 -\frac{1}{2} \wspt^{0,\mu\nu} \omega_{\mu\nu}(x) \right) 
\ee
for which it can be shown that, with calculations similar to those in the previous
section, the variation induced by the transformation (\ref{transfq}) is:
\be\label{dwmax}
 \delta \wUpsl =\frac{1}{2}\left\{\int \di S 
 \, n_i \, \wvphi^{i 0,\nu} \delta \beta_\nu - \int \di^3 \x  \left[  \wPhi^{\lambda, 0 \nu}
 (\partial_\lambda \delta \beta_\nu + \partial_\nu \delta \beta_\lambda) 
 - \wPhi^{0, \lambda \nu} \left( \frac{1}{2} (\partial_\lambda \delta \beta_\nu - 
 \partial_\nu \delta \beta_\lambda)  + \delta \omega_{\lambda \nu} \right) \right] \right\}
\ee
As has been mentioned, it is possible to get rid of the surface and the last term 
in the right hand side of above equation through a suitable choice of the perturbations, 
but not of the second term.

Since $\dwUps_{\rm LE}$ is a small term compared to $\wUps_{\rm LE}$ we can determine 
the variation of the entropy (\ref{ent}) with an expansion in $\dwUpsl$ at first 
order. First, we observe that (see also Eq.~(\ref{zass})):
$$
 Z'_{\rm LE} \equiv \tr \left(\exp[-\wUpsl-\dwUpsl]\right) \simeq 
 \tr \left(\exp[-\wUpsl](I-\dwUpsl)\right) = Z_{\rm LE}(1-\aveqU{\dwUpsl})
$$ 
where $\aveqU{\,\,}$ stands for the averaging with the original $\wUpsl$ local equilibrium
operator. Hence, the new entropy reads:
\bea\label{ent2}
 S' &=& \frac{1}{Z'_{\rm LE}} \tr \left(\exp[-\wUpsl-\dwUpsl] (\wUpsl +\dwUpsl)\right) 
 + \log Z'_{\rm LE} \nonumber \\
 &\simeq& \frac{1}{Z_{\rm LE}} (1+\aveqU{\dwUpsl}) \, \tr \left(\exp[-\wUpsl-\dwUpsl] 
 (\wUpsl +\dwUpsl)\right) + \log Z_{\rm LE} + \log(1-\aveqU{\dwUpsl}) 
\eea
We can now further expand the exponentials as we have done in Eq.~(\ref{zass}). First:
\bea
 && \tr \left(\exp[-\wUpsl - \dwUpsl] \wUpsl\right) \simeq \tr \left(\exp[-\wUpsl](I - 
 \dwUpsl + \frac{1}{2} [\wUpsl,\dwUpsl] - \frac{1}{6} [\wUpsl,[\wUpsl,\dwUpsl]] 
 + \ldots) \wUpsl\right) \nonumber \\
 && = \tr (\exp[-\wUpsl]\wUpsl) -  \tr (\exp[-\wUpsl]\dwUpsl \, \wUpsl) 
 = Z_{\rm LE} \aveqU{\wUpsl} - Z_{\rm LE} \aveqU{\dwUpsl\,\wUpsl}
\eea
where, in the second equality, we have taken advantage of commutativity and cyclicity
of the trace. Then: 
\bea 
 && \tr \left(\exp[-\wUpsl - \dwUpsl] \dwUpsl\right) \simeq \tr \left(\exp[-\wUpsl](I - 
 \dwUpsl + \frac{1}{2} [\wUpsl,\dwUpsl] - \frac{1}{6} [\wUpsl,[\wUpsl,\dwUpsl]] 
 + \ldots) \dwUpsl\right) \nonumber \\
 && \simeq \tr (\exp[-\wUpsl]\dwUpsl)= Z_{\rm LE} \aveqU{\dwUpsl}
\eea
keeping only first order terms. Thus, Eq.~(\ref{ent2}) can be rewritten as:
\bea\label{ent3}
 S' &\simeq& \frac{1}{Z_{\rm LE}} (1+\aveqU{\dwUpsl}) \, \tr \left(\exp[-\wUpsl-\dwUpsl] 
 (\wUpsl +\dwUpsl)\right) + \log Z_{\rm LE} + \log(1-\aveqU{\dwUpsl}) \nonumber \\
 &\simeq& \frac{1}{Z_{\rm LE}} (1+\aveqU{\dwUpsl}) 
 \left(Z_{\rm LE} \aveqU{\wUpsl} - Z_{\rm LE} \aveqU{\dwUpsl\,\wUpsl}
 + Z_{\rm LE} \aveqU{\dwUpsl}\right) + \log Z_{\rm LE} + \log(1-\aveqU{\dwUpsl}) 
 \nonumber\\
  &=& (1+\aveqU{\dwUpsl})\left(\aveqU{\wUpsl} - \aveqU{\dwUpsl\,\wUpsl} + 
  \aveqU{\dwUpsl} \right) + \log Z_{\rm LE} + \log(1-\aveqU{\dwUpsl}) 
\eea
Retaining only the first order terms in $\dwUpsl$, expanding the logarithm for
$\aveql{\dwUpsl} \ll 1$ and inserting the original expression of entropy:
\be\label{ent4}
  S' \simeq S -\aveqU{\dwUpsl\,\wUpsl} + \aveqU{\dwUpsl}\aveqU{\wUpsl}
\ee
Therefore, the variation of the total entropy is, to the lowest order, proportional
to the correlation between $\wUps$ and $\dwUps$, which is generally non-vanishing.

We can expand the above correlation to gain further insight. For the $\dwUpsl$, let 
us keep only the second term of the right hand side of Eq.~(\ref{dwmax}):
\be\label{dups}
 \dwUpsl = - \frac{1}{2} \int \di^3 \x  \; \wPhi^{\lambda, 0 \nu}
 (\partial_\lambda \delta \beta_\nu + \partial_\nu \delta \beta_\lambda) 
\ee
By using the (\ref{loceq}) and the (\ref{dups}), the Eq.~(\ref{ent4}) can be rewritten 
as: 
\bea\label{ent5}
  S'(t) \simeq S(t) &+& \frac{1}{2}\int \di^3 \x \int \di^3 \x' \; \left( 
  \aveqU{\wPhi^{\lambda, 0 \nu}(x) \, \wT^{0 \mu}(x')} - \aveqU{\wPhi^{\lambda, 0 \nu}(x)} 
  \aveqU{\wT^{0 \mu}(x')} \right) \beta_\mu(x') (\partial_\lambda \delta \beta_\nu (x) + 
  \partial_\nu \delta \beta_\lambda(x)) \nonumber \\
  &-& \frac{1}{2}\int \di^3 \x \int \di^3 \x' \; \left( \aveqU{\wPhi^{\lambda, 0 \nu}(x) 
  \, \wj^{0}(x')} - \aveqU{\wPhi^{\lambda, 0 \nu}(x)}\aveqU{\wj^{0}(x')} \right) \xi(x') 
  (\partial_\lambda \delta \beta_\nu (x) + \partial_\nu \delta \beta_\lambda(x)) \nonumber \\
  &-& \frac{1}{4}\int \di^3 \x \int \di^3 \x' \; \left( \aveqU{\wPhi^{\lambda, 0 \nu}(x) 
  \, \wspt^{0,\rho \sigma}(x')} - \aveqU{\wPhi^{\lambda, 0 \nu}(x)}\aveqU{\wspt^{0,\rho \sigma}(x')} 
  \right) \omega_{\rho \sigma}(x')
  (\partial_\lambda \delta \beta_\nu (x) + \partial_\nu \delta \beta_\lambda(x))
  \nonumber \\
\eea
where $x$ and $x'$ have equal times. The above expression could be further simplified
by e.g. approximating the local equilibrium mean $\aveqU{ \; }$ with the global 
equilibrium one $\aveq{ \; }$, but this does not lead to further conceptual insight.
The physical meaning of Eq.~(\ref{ent5}) is that the entropy difference depends on
the correlation between local operators in two different space points multiplied 
by a factor which is at most of the second order in the perturbation $\delta\beta$. 
This kind of expression resembles the product of transport coefficients expressed 
by a Kubo formula times the squared gradient of the perturbation field. Therefore, 
the difference between entropies suggest that the introduction of a superpotential 
may lead to a modification of the transport coefficients. We will show this in detail 
in the next Section. 

\section{Transport coefficients: shear viscosity as an example}
\label{kubo}

As has been mentioned, a remarkable consequence of the transformation (\ref{transfq}) 
is a difference in the predicted values of transport coefficients calculated with 
the relativistic Kubo 
formula, which is obtained by working out the mean value of the stress-energy tensor 
itself with the linear response theory and the nonequilibrium density operator in 
Eq.~(\ref{zub1}). For this purpose, the derivation in ref.~\cite{hosoya} must be 
extended to the most general expression of the nonequilibrium density operator 
including a spin tensor, that is, Eq.~(\ref{zub2}); it can be found in Appendix A.

The equation (\ref{lrto}), yielding the difference of mean values of a general observable
under a transformation (\ref{transfq}), cannot be straightforwardly used to calculate
the mean value of the stress-energy tensor setting $\widehat O = \wT^{\mu \nu}(y)$ 
because $\wT^{\mu \nu}(y)$ gets transformed itself. It is therefore more convenient 
to work out the general expression of the Kubo formula and study how it is modified 
by (\ref{transfq}) thereafter.

We will take shear viscosity as an example, the transformation of other transport 
coefficients can be obtained with the same reasoning. Shear viscosity, in the Kubo
formula, is related to the spatial components of the symmetric part of the 
stress-energy tensor. It is worth pointing out that, since a non-vanishing spin tensor 
can make the stress-energy tensor non-symmetric, there might be a new transport 
coefficient related to the antisymmetric part of the stress-energy tensor.

For the symmetric part of the stress-energy tensor $T_S^{\mu\nu} \equiv (1/2) (T^{\mu \nu}
+ T^{\nu \mu})$, using the general formula of relativistic linear response theory 
(Eq. \ref{meanobsf}) of Appendix A), the difference $\delta T_S^{\mu\nu}(y)$ between 
actual mean value and local equilibrium value reads, at the lowest order in gradients:
\bea\label{deltt1}
 \delta T_S^{\mu\nu}(y) &=& 
 \lim_{\varepsilon\to 0} \; \frac{T}{\ii} \int_{-\infty}^{t'} \di t \;
 \frac{1-\ex{\varepsilon(t-t')}}{\varepsilon} \int \di^3 \x \;  
 \aveq{\comm{\wT^{\rho\sigma}(x),\wT_S^{\mu\nu}(y)}}\partial_\rho\deltabeta_\sigma(x) \nonumber \\
 && - \frac{1}{2} \lim_{\varepsilon\to 0} \frac{T}{\ii} \int_{-\infty}^{t'} \di t \;
 \ex{\varepsilon(t-t')} \int \di^3 \x \; 
 \aveq{\comm{\wspt^{0,\rho\sigma}(x),\wT_S^{\mu\nu}(y)}} \delta\omega_{\rho\sigma}(x) \nonumber \\
 && - \frac{1}{2} \lim_{\varepsilon\to 0} \frac{T}{\ii} \int_{-\infty}^{t'} \di t \;
 \ex{\varepsilon(t-t')}  \int_{-\infty}^{t} \di \tau \int \di^3 \x \; 
 \aveq{\comm{\wspt^{0,\rho\sigma}(\tau,{\bf x}),\wT_S^{\mu\nu}(y)}} \frac{\partial}{\partial t}
 \delta\omega_{\rho\sigma}(x)
\eea
In order to obtain transport coefficients, a suitable perturbation must be chosen 
which can be eventually taken out from the integral. Physically, this corresponds to 
enforcing a particular hydrodynamical motion and observing the response of the 
stress-energy tensor to infer the dissipative coefficient. The perturbation 
$\delta \beta = 1/T \delta u$  is taken to be a stationary one and non-vanishing 
only within a finite region $V$, at whose boundary it goes to zero in a continuous 
and derivable fashion. 
The perturbation $\delta\omega$ is also taken to be stationary and it can be chosen 
either to vanish or like in Eq.~(\ref{ombetalock}); in both cases, one gets to the 
same final result. 

Let us then set $\delta\omega=0$ and expand the perturbation $\delta\beta = 
(0,0,\delta\beta^2(x^1),0)$ dependent on $x^1$ in a Fourier series (it vanishes at some 
large, yet finite boundary). Since we want the higher order gradients of the 
perturbation to be negligibly small (the so-called hydrodynamic limit), the 
Fourier components with short wavelengths must be correspondingly suppressed. 
The component with the longest wavelength will then be much larger than any 
other and, therefore, $\delta \beta^2$ can be approximately written, 
at least far from the boundary, as $A \sin(\pi x^1/L)$ where $L$ is the size of 
the region $V$ in the $x^1$ direction and $A$ is a constant. The derivative of 
this perturbation reads:
$$
 \partial_1\deltabeta_2 ({\bf x}) = \frac{\pi}{L} A \cos(\pi x^1/L) =
 \partial_1\deltabeta_2 ({\bf 0}) \cos(\pi x^1/L) \equiv 
 \partial_1\deltabeta_2 ({\bf 0}) \cos(k x^1)
$$
where $k \equiv \pi/L$. Therefore, by defining ${\bf k} = (k,0,0)$ and plugging 
the last equation in Eq.~(\ref{deltt1}):
\bea\label{deltt2}
 \delta T_S^{\mu\nu}(y) &=& 
 \lim_{\varepsilon\to 0} \; \frac{T}{\ii} \partial_1\deltabeta_2 ({\bf 0})
 \int_{-\infty}^{t'} \di t \;\frac{1-\ex{\varepsilon(t-t')}}{\varepsilon} 
 \int_V \di^3 \x \; \cos{{\bf k}\cdot{\bf x}} \aveq{\comm{\wT^{12}(x),\wT_S^{\mu\nu}(y)}}
 \nonumber \\
 &=& \lim_{\varepsilon\to 0} \; T \, \partial_1\deltabeta_2 ({\bf 0}) \;
 {\rm Im} \int_{-\infty}^{t'} \di t \;\frac{1-\ex{\varepsilon(t-t')}}{\varepsilon} 
 \int_V \di^3 \x \; \e^{\ii{\bf k}\cdot{\bf x}} \aveq{\comm{\wT^{12}(x),\wT_S^{\mu\nu}(y)}}
\eea
taking into account that the commutator is purely imaginary. To extract shear viscosity
we have to evaluate the stress-energy tensor in ${\bf y}=0$ to make it proportional
to the derivative of the four-temperature field in the same point and we have to 
take the limit $L\to \infty$ which implies $V \to \infty$ and ${\bf k} \to 0$ at 
the same time:
\be\label{deltt3}
 \delta T_S^{\mu\nu}(t_y,{\bf 0}) = 
 \lim_{\varepsilon\to 0} \lim_{{\bf k}\to 0}\;T \, 
 \partial_1\deltabeta_2 ({\bf 0}) \; {\rm Im} \int_{-\infty}^{t'} \di t 
 \;\frac{1-\ex{\varepsilon(t-t')}}{\varepsilon} \int \di^3 \x \; 
 \e^{\ii{\bf k}\cdot{\bf x}} \aveq{\comm{\wT^{12}(x),\wT_S^{\mu\nu}(t_y,{\bf 0})}}
\ee
where it has been assumed that the integration domain goes to its thermodynamic
limit independently of the integrand.
Because of the time-translation symmetry of the equilibrium density operator $\wrho_0$,
the mean value in the integral only depends on the time difference $t-t_y$. Thus,
choosing the arbitrary time $t'=t_y$ and redefining the integration variables, 
the Eq.~(\ref{deltt3}) can be rewritten as:
\be\label{deltt4}
 \delta T_S^{\mu\nu}(t_y,{\bf 0}) = 
 \lim_{\varepsilon\to 0} \lim_{{\bf k}\to 0}\; T \,
 \partial_1\deltabeta_2 ({\bf 0}) \; {\rm Im} \int_{-\infty}^{0} \di t 
 \;\frac{1-\ex{\varepsilon t}}{\varepsilon} \int \di^3 \x \; 
 \e^{\ii{\bf k}\cdot{\bf x}} \aveq{\comm{\wT^{12}(x),\wT_S^{\mu\nu}(0)}}
\ee
which shows that the mean value $\delta T_S^{\mu\nu}(t_y,{\bf 0})$ is indeed 
independent of $t_y$, which is expected as $\delta\beta$ is stationary.
 
We can now take advantage of the well known Curie symmetry ``principle" which states 
that tensors belonging to some irreducible representation of the rotation group will 
only respond to perturbations belonging to the same representation and with the
same components \footnote{This is true provided that the right hand side of Eq.~(\ref{deltt4})
is a continuous function of ${\bf k}$ for ${\bf k}=0$ or that its limit for 
${\bf k}\to 0$ exists, i.e. it is independent of the direction of ${\bf k}$}. 
In our case the Curie principle implies that only the same component of the 
symmetric part of the stress-energy tensor, i.e. $\wT_S^{12}$, will give a 
non-vanishing value:
\be\label{deltt5}
 \delta T_S^{12}(t_y,{\bf 0}) = 
 \lim_{\varepsilon\to 0} \lim_{{\bf k}\to 0}\; T \,
 \partial_1\deltabeta_2 ({\bf 0}) \; {\rm Im} \int_{-\infty}^{0} \di t 
 \;\frac{1-\ex{\varepsilon t}}{\varepsilon} \int \di^3 \x \; 
 \e^{\ii{\bf k}\cdot{\bf x}} \aveq{\comm{\wT_S^{12}(x),\wT_S^{12}(0)}}
\ee
From the above expression, a Kubo formula for shear viscosity can be extracted 
setting $\delta \beta = (1/T) \delta u$: 
\be\label{visco}
 \eta = \lim_{\varepsilon\to 0} \lim_{{\bf k} \to 0} \; {\rm Im} 
 \int_{-\infty}^0 \di t \; \frac{1-\e^{\varepsilon t}}{\varepsilon} 
 \int \di^3 \x \; \e^{\ii {\bf k} \cdot {\bf x}} 
 \aveq{\comm{\wT_S^{12}(x),\wT_S^{12}(0)}}
\ee  
which, after a little algebra, can be shown to be the same expression obtained in 
ref.~\cite{hosoya}. Because of the rotational invariance of the equilibrium density
operator, shear viscosity is independent of the particular couple $(1,2)$ of chosen 
indices. It is worth pointing out that, had we started from Eq.~(\ref{meanobsf2})
instead of Eq.~(\ref{meanobsf}), choosing $\delta\omega=0$ or like in Eq.~(\ref{ombetalock}), 
we would have come to the same formula for shear viscosity; in the latter case, 
the third contributing term in Eq.~(\ref{meanobsf2}) would have been of higher 
order in derivatives of $\delta\beta$, hence negligible.

Now, the question we want to answer is whether equation (\ref{visco}) is invariant by 
a pseudo-gauge transformation (\ref{transfq}), which turns the symmetric part of the
stress-energy tensor into:
\be\label{tvar}
  \wT'^{\mu\nu}_S =  \wT^{\mu\nu}_S - \frac{1}{2}
   \partial_\lambda (\wPhi^{\mu,\lambda \nu} + \wPhi^{\nu, \lambda \mu}) 
  = \wT^{\mu\nu}_S - \partial_\lambda \wXi^{\lambda \mu \nu}
\ee
where:
\be\label{xite}
 \frac{1}{2} (\wPhi^{\mu,\lambda \nu} + \wPhi^{\nu, \lambda \mu}) 
 \equiv \wXi^{\lambda \mu \nu}
\ee
$\wXi$ being symmetric in the last two indicess. We will study the effect of the 
transformation on the mean value of the stress-energy tensor in the point $y=0$ starting 
from the formula Eq.~(\ref{meanobsf2}) instead of Eq.~(\ref{meanobsf}) with 
$\delta\omega=0$ or like in Eq.~(\ref{ombetalock}), which allows us to retain only
the first contributing term to $\delta T_S^{12}(0)$. The perturbation $\delta\beta$
is taken to be stationary and $t'$ is set to be equal to $t_y = 0$. Eventually,
the appropriate limits will be calculated to get the new shear viscosity. Thus:
\bea\label{visct1}
&& \delta T_S^{'12}(0) = \delta T_S^{12}(0) 
 + \lim_{\varepsilon\to 0}\; \int_{-\infty}^0 \di t \; 
 \frac{1-\ex{\varepsilon t}}{\varepsilon} \int \di^3 \x \; 
 \aveq{\comm{\partial_\alpha \wXi^{\alpha 12}(x),\partial_\beta \wXi^
 {\beta 12}(0)}}(\partial_1\deltabeta_2({\bf x})+\partial_2\deltabeta_1({\bf x})) 
\\ \nonumber
&& - \lim_{\varepsilon\to 0} \int_{-\infty}^0 \di t \; 
 \frac{1-\ex{\varepsilon t}}{\varepsilon} \int \di^3 \x \; \left( 
 \aveq{\comm{\partial_\alpha\wXi^{\alpha 12}(x),\wT^{12}_{S}(0)}} 
 + \aveq{\comm{\wT^{12}_{S}(t,{\bf x}),\partial_\alpha\wXi^{\alpha 12}(0)}}
 \right)(\partial_1\deltabeta_2({\bf x})+\partial_2\deltabeta_1({\bf x})) 
\eea
We can simplify the above formula 
by noting that the mean value of two operators at equilibrium can oly depend on the 
difference of the coordinates, so:
$$
 \aveq{\comm{\widehat O_1(y),\partial_\mu\widehat O_2(x)}} = 
 \frac{\partial}{\partial x^\mu}\aveq{\comm{\widehat O_1,\widehat O_2}}(y-x) =  
 -\frac{\partial}{\partial y^\mu}\aveq{\comm{\widehat O_1,\widehat O_2}}(y-x),
$$
hence, the Eq.~(\ref{visct1}) can be rewritten as:
\bea\label{visct2}
&& \delta T_S^{'12}(0) = \delta T_S^{12}(0) 
 - \lim_{\varepsilon\to 0} \int_{-\infty}^0 
 \di t \; \frac{1-\ex{\varepsilon t}}{\varepsilon} \int \di^3 \x \; 
 \frac{\partial^2}{\partial x^\alpha \partial x^\beta} 
 \aveq{\comm{\wXi^{\alpha 12}(x),\wXi^{\beta 12}(0)}}
 (\partial_1\deltabeta_2({\bf x})+\partial_2\deltabeta_1({\bf x})) \nonumber \\
&& - \lim_{\varepsilon\to 0} \int_{-\infty}^0 \di t \; 
 \frac{1-\ex{\varepsilon t}}{\varepsilon} \int \di^3 \x \; \frac{\partial}{\partial x^\alpha}
 \left(\aveq{\comm{\wXi^{\alpha 12}(x),\wT^{12}_{S}(0)}} 
 - \aveq{\comm{\wT^{12}_{S}(x),\wXi^{\alpha 12}(0)}}
 \right)(\partial_1\deltabeta_2({\bf x})+\partial_2\deltabeta_1({\bf x}))
\eea
We are now going to inspect the two terms on the right-hand side of the above equation.
If the hamiltonian is time-reversal invariant, it can be shown (see Appendix B):
$$
 \aveq{\comm{\wT^{ij}_S(t,{\bf x}),\wXi^{\alpha ij}(0,{\bf 0})}} = 
  (-1)^{n_0} \aveq{\comm{\wXi^{\alpha ij}(0,{\bf 0}),\wT^{ij}_S(-t,{\bf x})}} 
= (-1)^{n_0} \aveq{\comm{\wXi^{\alpha ij}(t,-{\bf x}),\wT^{ij}_S(0,{\bf 0})}}
$$
where $n_0$ is the total number of time indices among those in the above expression. 
Similarly, if the hamiltonian is parity invariant, then:
$$
 \aveq{\comm{\wXi^{\alpha ij}(t,-{\bf x}),\wT^{ij}_S(0,{\bf 0})}} 
 = (-1)^{n_s} \aveq{\comm{\wXi^{\alpha ij}(t,{\bf x}),\wT^{ij}_S(0,{\bf 0})}}
$$
where $n_s$ is the total number of space indices. Using the last two equations
to work out the last term of Eq.~(\ref{visct2}) one gets:
\bea\label{visct3}
&& \delta T_S^{'12}(0) = \delta T_S^{12}(0) 
 - \lim_{\varepsilon\to 0} \int_{-\infty}^0 
 \di t \; \frac{1-\ex{\varepsilon t}}{\varepsilon} \int_V \di^3 \x \; 
 \frac{\partial^2}{\partial x^\alpha \partial x^\beta} 
 \aveq{\comm{\wXi^{\alpha 12}(t,{\bf x}),\wXi^{\beta 12}(0,{\bf 0})}}
 (\partial_1\deltabeta_2({\bf x})+\partial_2\deltabeta_1({\bf x})) \nonumber \\
&& - 2 \lim_{\varepsilon\to 0} \int_{-\infty}^0 \di t \; 
 \frac{1-\ex{\varepsilon t}}{\varepsilon} \int_V \di^3 \x \; 
  \frac{\partial}{\partial x^\alpha}
 \aveq{\comm{\wXi^{\alpha 12}(t,{\bf x}),\wT^{12}_{S}(0,{\bf 0})}} 
 (\partial_1\deltabeta_2({\bf x})+\partial_2\deltabeta_1({\bf x})) 
\eea
Now, the two terms on the right hand side of (\ref{visct3}) can be worked out 
separately. Using invariance by time-reversal and parity, one has:
\bea
 && \aveq{\comm{\wXi^{\alpha ij}(t,{\bf x}),\wXi^{\beta ij}(0,{\bf 0})}} = 
  (-1)^{n_0}\aveq{\comm{\wXi^{\beta ij}(0,{\bf 0}),\wXi^{\alpha ij}(-t,{\bf x})}} 
  \nonumber \\
 &=& (-1)^{n_0}\aveq{\comm{\wXi^{\beta ij}(t,-{\bf x}),\wXi^{\alpha ij}(0,{\bf 0})}}
 = (-1)^{n_0+n_s}\aveq{\comm{\wXi^{\beta ij}(t,{\bf x}),\wXi^{\alpha ij}(0,{\bf 0})}}
 = \aveq{\comm{\wXi^{\beta ij}(t,{\bf x}),\wXi^{\alpha ij}(0,{\bf 0})}}
\eea
being $n_0+n_s=6$. Hence, the first term on the right hand side of (\ref{visct3}) 
can be decomposed as:
\bea\label{first}
 && - \lim_{\varepsilon\to 0} \int_{-\infty}^0\di t  \frac{1-\ex{\varepsilon t}}{\varepsilon} 
 \int_V \di^3 \x \; \left( \frac{\partial^2}{\partial t^2} 
 \aveq{\comm{\wXi^{0 ij}(x),\wXi^{0 ij}(0)}} + 2 \frac{\partial}{\partial t} 
 \frac{\partial}{\partial x^k}\aveq{\comm{\widehat \Xi^{k ij}(x),\widehat \Xi ^{0 ij}(0)}}  
 \right. \nonumber \\ 
 && \left. + \frac{\partial}{\partial x^k}\frac{\partial}{\partial x^l}
 \aveq{\comm{\wXi^{k ij}(x),\wXi^{l ij}(0)}} \right)
 (\partial_i\deltabeta_j({\bf x}) + \partial_j\deltabeta_i({\bf x})) 
\eea
and, similarly, the second term as:
\be\label{second}
 - 2 \lim_{\varepsilon\to 0} \int_{-\infty}^0 \di t \; 
 \frac{1-\ex{\varepsilon t}}{\varepsilon} \int_V \di^3 \x \; 
 \frac{\partial}{\partial t} \aveq{\comm{\wXi^{0 12}(x),\wT^{12}_{S}(0)}} 
+ \frac{\partial}{\partial x^k} \aveq{\comm{\wXi^{k12}(t,{\bf x}),\wT^{12}_{S}(0)}}
 (\partial_1\deltabeta_2({\bf x})+\partial_2\deltabeta_1({\bf x})) 
\ee
All terms in Eqs.~(\ref{first}) and (\ref{second}) with a space derivative do not 
yield any contribution to first-order transport coefficients. This can be shown by, 
firstly, integrating by parts and generating two terms, one of which is a total 
derivative and the second involves the second derivative of the perturbation $\delta \beta$. 
The total derivative term can be transformed into a surface integral on the boundary 
of $V$ which vanishes because therein the perturbation $\delta \beta$ is supposed 
to vanish along with its first-order derivatives. The second term, involving higher 
order derivatives, does not give contribution to transport coefficients at first 
order in the derivative expansion. Altogether, the Eq.~(\ref{visct3}) turns into:
\bea\label{visct4}
&& \delta T_S^{'12}(0) = \delta T_S^{12}(0) - \lim_{\varepsilon\to 0} \int_{-\infty}^0 
 \di t \; \frac{1-\ex{\varepsilon t}}{\varepsilon} \int_V \di^3 \x \; 
 \partial^2_t \aveq{\comm{\wXi^{0 12}(x),\wXi^{0 12}(0)}}
 (\partial_1\deltabeta_2({\bf x})+\partial_2\deltabeta_1({\bf x})) \nonumber \\
&& - 2 \lim_{\varepsilon\to 0} \int_{-\infty}^0 \di t \; 
 \frac{1-\ex{\varepsilon t}}{\varepsilon} \int_V \di^3 \x \; 
  \partial_t \aveq{\comm{\wXi^{012}(x),\wT^{12}_{S}(0)}} 
 (\partial_1\deltabeta_2({\bf x})+\partial_2\deltabeta_1({\bf x})) + 
 {\cal O}(\partial^2 \delta\beta)
\eea
which can be further integrated by parts in the time $t$, yielding:
\bea\label{visct5}
&& \delta T_S^{'12}(0) = \delta T_S^{12}(0) - \lim_{\varepsilon\to 0} \int_{-\infty}^0 
 \di t \; (\delta(t) - \varepsilon \, \e^{\varepsilon t}) \int_V \di^3 \x \; 
 \aveq{\comm{\wXi^{0 12}(x),\wXi^{0 12}(0)}}
 (\partial_1\deltabeta_2({\bf x})+\partial_2\deltabeta_1({\bf x})) \nonumber \\
&& - 2 \lim_{\varepsilon\to 0} \int_{-\infty}^0 \di t \; 
  \ex{\varepsilon t} \int_V \di^3 \x \; \aveq{\comm{\wXi^{012}(x),\wT^{12}_{S}(0)}} 
 (\partial_1\deltabeta_2({\bf x})+\partial_2\deltabeta_1({\bf x})) + 
 {\cal O}(\partial^2 \delta\beta)
\eea
provided that, for general space-time dependent operators $\widehat O_1$ and $\widehat O_2$
$$
 \lim_{t \to -\infty} \int_V \di^3 \x \; \e^{n\varepsilon t}
  \frac{\partial}{\partial t} \aveq{\comm{\widehat O_1(t,{\bf x}),
  \widehat O_2(0,{\bf 0})}} = 0
$$
$$  
  \lim_{t \to -\infty} \int_V \di^3 \x \; \e^{n\varepsilon t}
  \aveq{\comm{\widehat O_1(t,{\bf x}),\widehat O_2(0,{\bf 0})}} = 0
$$
with $n=0,1$, which is reasonable because thermodynamical correlations are expected
to vanish exponentially as a function of time for fixed points in space 
\footnote{\label{lcone} There might be singularities on the light cone, however for fixed {\bf x} 
and {\bf 0} and integration over a finite region $V$, in the limit $t \to -\infty$ 
light cone is not involved}. 

From Eq.~(\ref{visct5}) the variation of the shear viscosity can be inferred with 
the very same reasoning that led us to formula (\ref{visco}), that is:
\bea\label{deltaeta}
\Delta \eta = \eta' - \eta = &-& \lim_{\varepsilon\to 0} \lim_{k \to 0}\; {\rm Im}  
 \int_{-\infty}^0 \di t \; (\delta(t) - \varepsilon \, \e^{\varepsilon t}) 
  \int \di^3 \x \; \e^{\ii k x^1} \aveq{\comm{\wXi^{0 12}(t,{\bf x}),\wXi^{0 12}(0,{\bf 0})}}
   \nonumber \\
&-& 2 \lim_{\varepsilon\to 0} \lim_{k \to 0}\; {\rm Im} \int_{-\infty}^0 \di t \; 
  \ex{\varepsilon t} \int \di^3 \x \; \e^{\ii k x^1} 
   \aveq{\comm{\wXi^{012}(t,{\bf x}),\wT^{12}_{S}(0,{\bf 0})}} 
\eea
If the first integral is regular, then the $\varepsilon \to 0$ limit kills one term
and the (\ref{deltaeta}) reduces to:
\bea\label{deltaeta2}
\Delta \eta = \eta' - \eta = &-& \lim_{k \to 0}\; 
  \int_V \di^3 \x \; \cos k x^1 \aveq{\comm{\wXi^{0 12}(0,{\bf x}),\wXi^{0 12}(0,{\bf 0})}}
   \nonumber \\
&-& 2 \lim_{\varepsilon \to 0} \lim_{k \to 0}\; {\rm Im} \int_{-\infty}^0 \di t \; 
  \ex{\varepsilon t} \int \di^3 \x \; \e^{\ii k x^1} 
   \aveq{\comm{\wXi^{012}(x),\wT^{12}_{S}(0,{\bf 0})}} 
\eea
In general, this difference is non-vanishing, leading to the conclusion that the 
specific form of the stress-energy tensor and, possibly, the existence of a spin 
tensor in the underlying quantum field theory affects the value of transport 
coefficients. The relative difference of those values depends on the particular 
transformation (\ref{transfq}), hence on the particular stress-energy tensor. In 
the next Section a specific instance will be presented and discussed.

An important point to make is that the found dependence of the transport coefficients on 
the particular set of stress-energy and spin tensor of the theory is indeed physically 
meaningful. This means that the variation of some coefficient is not compensated by a 
corresponding variation of another coefficient so as to eventually leave measurable 
quantities unchanged. This has been implicitely proved in Sect.~\ref{mean} where it
was shown that total entropy itself undergoes a variation under a transformation of 
the stress-energy and spin tensor (see Eq.~(\ref{ent4})).

\section{Discussion and conclusions}
\label{discu}

As a first point, we would like to emphasize that in our arguments space-time curvature 
and gravitational coupling have been disregarded. On one hand, this shows that 
the nature of stress-energy tensor and, possibly, the existence of a fundamental 
spin tensor could, at least in principle, be demonstrated independently of gravity. 
On the other hand, for each stress-energy tensor created with the transformation 
(\ref{transfq}), it should be shown that an extension of general relativity exists 
having it as a source, which could not be always possible.

An important question is whether a concrete physical system indeed exists for which the 
transformation (\ref{transfq}) leads to actually different values for e.g. transport 
coefficients, entropy production rate or other quantities in nonequilibrium situations. 
For this purpose, we discuss a specific instance regarding spinor electrodynamics. 
Starting from the symmetrized gauge-invariant Belinfante tensor of the coupled Dirac 
and electromagnetic fields, with associated $\wspt=0$:
\be\label{belinf}  
   \wT^{\mu\nu} = \frac{\ii}{4} \left( \Psibar \gamma^\mu \codevc^\nu \Psi 
   + \Psibar \gamma^\nu \codevc^\mu \Psi \right) + \widehat F^{\mu}_{\;\;\lambda} 
   \widehat F^{\lambda \nu} + \frac{1}{4} g^{\mu\nu} \widehat F^2    
\ee
where $\nabla_\mu = \partial_\mu - \ii e A_\mu$ is the gauge covariant derivative, one
can generate other stress-energy tensors with suitable rank three tensors and then 
setting $\wPhi=-\wspt'$ where $\wspt'$ is the new spin tensor, according to (\ref{transfq}). 
One of the best known is the {\em canonical} Dirac spin tensor:
$$
 \wPhi^{\lambda,\mu\nu} = - \frac{i}{8}\Psibar \{\gamma^\lambda, [\gamma^\mu,\gamma^\nu] \} \Psi
$$
($\{ \; \}$ stands for anticommutator) which is gauge-invariant and transforms the 
Belinfante tensor (\ref{belinf}) back to the canonical one obtained from the spinor 
electrodynamics lagrangian (see also \cite{bt1} for a detailed discussion). However, 
this is totally antysimmetric in the three indices $\lambda,\mu,\nu$ and thus the 
variation of $\wUps$ operator (see Eq.~(\ref{deltaups}) as well as transport 
coefficients, which depend on the symmetrized $\wXi$ tensor (\ref{xite}) vanish.     
Nevertheless, other gauge-invariant $\wPhi$-like tensors can be found. For instance, 
one could employ a superpotential:
$$
 \wPhi^{\lambda,\mu\nu} = \frac{1}{8 m} \Psibar \left( \gamma^\mu \codevc^\nu - \gamma^\nu
 \codevc^\mu \right) \gamma^\lambda \Psi + {\rm h.c} = 
 \frac{1}{8 m} \Psibar \left( [\gamma^\mu,\gamma^\lambda] \codevc^\nu 
  - [\gamma^\nu,\gamma^\lambda] \codevc^\mu  \right) \Psi
$$  
which is the gauge-invariant version of the one used in ref.~\cite{degroot} to obtain
a conserved spin current. This superpotential gives rise to a non-vanishing spin tensor
as well as a $\wXi$ tensor (see Eq.~\ref{xite})):
$$
 \wXi^{\lambda\mu\nu} = \frac{1}{16 m} 
 \Psibar \left( [\gamma^\lambda,\gamma^\mu] \codevc^\nu + [\gamma^\lambda,\gamma^\nu] 
 \codevc^\mu  \right) \Psi
$$  
hence a variation of thermodynamics. By noting that the structure of the above 
tensor is very similar to the Belinfante stress-energy tensor (\ref{belinf}), it is
not difficult to find a rough estimate of the variation of e.g. shear viscosity 
induced by the transformation. Looking at Eq.~(\ref{deltaeta2}) we note that 
$\wXi^{012}$ mainly differs from $\wT^{012}$ in Eq.~(\ref{belinf}) by the factor 
$1/m$. The last term on the right hand side of Eq.~(\ref{belinf}) tells us that
the dimension of$\wXi$ is that of a stress-energy tensor multiplied by a time, 
and therefore this term must be of the order of $\eta \hbar/m c^2 \tau$
where $\tau$ is the microscopic correlation time scale of the original stress-energy 
tensor or the collisional time scale in the kinetic language and $\eta$ the shear
viscosity obtained from the original stress-energy tensor. Thus, the expected
relative variation of shear viscosity from Eq.~(\ref{deltaeta2}) in this case 
is of the order:
$$
  \frac{\Delta \eta}{\eta} \approx {\cal O}\left( \frac{\hbar}{m c^2 \tau}\right)
$$
which is (as it could have been expected) a quantum relativistic correction 
governed by the ratio $(\lambda_c/c)/\tau$, $\lambda_c$ being the Compton wavelength. 
For the electron, the ratio $\lambda_c/c \approx 10^{-21}$ sec, which is a very 
small time scale compared to the usual kinetic time scales, yet it could be 
detectable for particular systems with very low shear viscosity.

It is also interesting to note that the ``improved" 
stress-energy tensor by Callan, Coleman and Jackiw \cite{callan} with renormalizable
matrix elements at all orders of perturbation theory, is obtained from the Belinfante's 
symmetrized one in Eq.~(\ref{belinf}) with a transformation of the kind (\ref{transfq})
setting (for the Dirac field and vanishing constants \cite{callan}):
$$
  \widehat Z^{\alpha\lambda,\mu\nu} = -\frac{1}{6} 
  \left(g^{\alpha \mu} g^{\lambda \nu} - g^{\alpha \nu} g^{\lambda \mu} \right) 
  \Psibar\Psi
$$
and requiring $\wspt'=\wspt=0$ so that $\wPhi^{\lambda,\mu\nu}= \partial_\alpha 
\widehat Z^{\alpha\lambda,\mu\nu}$, hence:
\begin{eqnarray*}
&&  \wPhi^{\lambda,\mu\nu} = -\frac{1}{6} \left( g^{\lambda \nu}\partial^\mu - 
     g^{\lambda \mu} \partial^\nu \right) \Psibar\Psi  \nonumber \\
&&  \wXi^{\lambda\mu\nu} = \frac{1}{2}( \wPhi^{\mu,\lambda\nu} + \wPhi^{\nu,\lambda\mu}) =
     -\frac{1}{6} \left[ g^{\mu\nu}\partial^\lambda - \frac{1}{2} (g^{\lambda \nu}
   \partial^\mu + g^{\lambda \mu}\partial^\nu) \right] \Psibar\Psi   \nonumber \\
&&  \wT'^{\mu\nu} = \wT^{\mu\nu} - \partial_\lambda \wXi^{\lambda\mu\nu} = \wT^{\mu\nu}
   + \frac{1}{6}(g^{\mu\nu} \Box - \partial^{\mu}\partial^\nu ) \Psibar\Psi
\end{eqnarray*}
which is just the improved stress-energy tensor \cite{callan}. It is likely (to be 
verified though) that the aforementioned modified stress-energy tensors imply a 
different thermodynamics with respect to the original Belinfante symmetrized tensor. 
This problem has been recently pointed out in ref.~\cite{nakayama}.

To summarize, we have concluded that different quantum stress-energy tensors imply
different values of nonequilibrium thermodynamical quantities like transport coefficients 
and entropy production rate. This reinforces our previous similar conclusion 
concerning differences of momentum and angular momentum densities in rotational 
equilibrium \cite{bt1}. The existence of a fundamental spin tensor has, thus, an 
impact on the microscopic number of degrees of freedom and on how quickly macroscopic 
information is converted into microscopic. The difference of transport coefficients 
depends on the particular form of the tensors and in the examined case it scales like a quantum 
relativistic effect with $\hbar/c$. Therefore, at least in principle, it is possible 
to disprove a supposed stress-energy tensor with a suitably designed thermodynamical 
experiment.

\section*{Acknowledgments}

We are grateful to F. Bigazzi, F.~W.~Hehl and D. Seminara for useful discussions
and suggestions.



\appendix

\section*{APPENDIX A - Relativistic linear response theory with spin tensor}

We extend the relativistic linear response theory in the Zubarev's approach to the
case of a non-vanishing spin tensor. The (stationary) nonequilibrium density operator
is written in Eq.~(\ref{zub2}), with $\wUps$ expanded as in Eq.~(\ref{elabor2}).
As has been shown in Sect.~\ref{zuba}, at equilibrium, only the first term of the
$\wUps$ operator survives in Eq.~(\ref{elabor2}); therefore, one can rewrite that
equation using the perturbations $\delta \beta$, $\delta \xi$ and $\delta \omega$
which are defined as the difference between the actual value and their value at
thermodynamical equilibrium:
\bea\label{elabor3}
 \wUps &=& \int \di^3 \x \; \left( \wT^{0\nu} \beta_\nu(t',{\bf x})
 - \wj^0 \xi(t',{\bf x}) -\frac{1}{2} \wspt^{0,\mu\nu} \omega_{\mu\nu}(t',{\bf x}) \right) 
 \nonumber \\
 &+& \lim_{\varepsilon \to 0} \int^{t'}_{-\infty} \!\!\! \di t \; \e^{\varepsilon (t-t')} 
 \int  \di S \, n_i \left( \wT^{i \nu} \delta\beta_\nu(x) - \wj^i \delta\xi(x) - 
 \frac{1}{2} \wspt^{i,\mu\nu} \delta\omega_{\mu\nu}(x)\right) \nonumber \\
 &-& \frac{1}{2} \lim_{\varepsilon \to 0} \int^{t'}_{-\infty} \!\!\! \di t \int \di^3 \x \; 
 \e^{\varepsilon (t-t')} \left( \wT_S^{\mu \nu} (\partial_\mu \delta\beta_\nu(x)+\partial_\mu 
 \delta\beta_\nu(x)) + \wT_A^{\mu \nu} (\partial_\mu \delta\beta_\nu(x) - 
 \partial_\mu \delta\beta_\nu(x) + 2 \delta\omega_{\mu\nu}(x)) \right. \nonumber \\
 && \left. - \wspt^{\lambda,\mu\nu} \partial_\lambda \delta\omega_{\mu\nu}(x) - 
 2 \wj^\mu \partial_\mu \delta\xi(x) \right) 
\eea
where it is understood that $x=(t,{\bf x})$.\\
In fact, we will use a rearrangement of the right-hand-side expression which is more
convenient if one wants to work with an unspecified, yet small, $\delta\omega$. Therefore,
the above equation is rewritten as:
\bea\label{elabor4}
 \wUps &=& \int \di^3 \x \; \left( \wT^{0\nu} \beta_\nu(t',{\bf x})
 - \wj^0 \xi(t',{\bf x}) -\frac{1}{2} \wspt^{0,\mu\nu} \omega_{\mu\nu}(t',{\bf x}) \right) 
 \nonumber \\
 &-& \lim_{\varepsilon \to 0} \int^{t'}_{-\infty} \!\!\! \di t \;  \e^{\varepsilon (t-t')}
 \frac{\partial}{\partial t} \int \di^3 \x \; \left(\wT^{0 \nu} \delta\beta_\nu(x) 
 - \frac{1}{2} \wspt^{0,\mu\nu} \delta\omega_{\mu\nu}(x) - \wj^0 \delta\xi(x) \right)
\eea
what it can be easily obtained from Eq.~(\ref{zub2}) integrating by parts in time. 

For the sake of simplicity we calculate the linear response with $\xi_{\rm eq}=
\delta\xi=0$, but it can be shown that our final expressions hold for $\xi_{\rm eq}
\ne0$ (in other words with a non-vanishing chemical potential $\mu \ne 0$). Let
us now define:
$$
 \widehat A = - \int \di^3 \x \; \left( \wT^{0\nu} \beta_\nu(t',{\bf x})
 - \frac{1}{2} \wspt^{0,\mu\nu} \omega_{\mu\nu}(t',{\bf x}) \right) 
$$ 
and:
$$
 \widehat B = \lim_{\varepsilon \to 0} \int^{t'}_{-\infty} \!\!\! \di t \; 
 \e^{\varepsilon (t-t')}\frac{\partial}{\partial t} \int \di^3 \x \; \left(\wT^{0 \nu} 
 \delta\beta_\nu(x) - \frac{1}{2} \wspt^{0,\mu\nu} \delta\omega_{\mu\nu}(x) \right)
$$
so that:
\be
 \wrho = \frac{1}{Z} \exp[-\wUps] = \frac{1}{Z} \exp[\widehat A + \widehat B]
\ee
with $Z=\tr (\exp[\widehat A + \widehat B])$. 

The operator $\widehat B$ is the small term in which $\wrho$ is to be expanded, 
according to the linear response theory. It can can be rewritten in a way which 
will be useful later on. Since:
\begin{eqnarray*}
   \int \di^3 \x \; \frac{\partial}{\partial t} \left( \wT^{0\nu}(x)\; \deltabeta_\nu(x) \right)
   &=& \int \di^3 \x \; \partial_\mu \left( \wT^{\mu\nu}(x) \, \deltabeta_\nu(x) \right) - 
   \int \di^3 \x \; \partial_i \wT^{i\nu}(x) \, \deltabeta_\nu(x) = \nonumber \\
&=& \int \di^3 \x \;  \wT^{\mu\nu}(x) \partial_\mu \deltabeta_\nu(x) - 
 \int_{\partial V} \di S \; {\hat n}_i \wT^{i\nu}(x) \, \deltabeta_\nu(x)
\end{eqnarray*}
then:
\begin{equation*}
 \widehat B = \lim_{\varepsilon \to 0} \int^{t'}_{-\infty} \!\!\! \di t \;
  \e^{\varepsilon (t-t')} \int \di^3 \x \; \left( \wT^{\mu \nu} \partial_\mu \delta\beta_\nu(x) 
 - \frac{1}{2} \frac{\partial}{\partial t} \left( \wspt^{0,\mu\nu} 
 \delta\omega_{\mu\nu}(x) \right) \right) - \int_{\partial V} \di S \; 
 {\hat n}_i \wT^{i\nu}(x) \, \deltabeta_\nu(x)
\end{equation*}
The perturbation $\delta\beta$ must be chosen such that $\deltabeta|_{\partial V}=0$
so that only the bulk term survives in the above equation:

\be\label{bterms}
 \widehat B = \lim_{\varepsilon \to 0} \int^{t'}_{-\infty} \!\!\! \di t \;
  \e^{\varepsilon (t-t')} \int \di^3 \x \; \left( \wT^{\mu \nu} \partial_\mu \delta\beta_\nu(x) 
 - \frac{1}{2} \frac{\partial}{\partial t} \left( \wspt^{0,\mu\nu} 
 \delta\omega_{\mu\nu}(x) \right) \right)
\ee

At the lowest order in $\widehat B$:
\be\label{expa1}
 Z =  \tr(\ex{\widehat A + \widehat B}) \simeq \tr(\ex{\widehat A}\comm{1+\widehat B}) 
 = Z_{\rm LE}(1 + \aveql{\widehat B}) \Rightarrow \frac{1}{Z}
 \simeq \frac{1}{Z_{\rm LE}}(1 - \aveql{\widehat B})
\ee
and, according to Kubo identity:
\be\label{expa2}
 \ex{\widehat A + \widehat B} = \comm{1+\int_0^1 \di z \; 
 \ex{z\brackets{\widehat A + \widehat B}}\widehat B \ex{-z\widehat A}}
 \ex{\widehat A}\simeq  \comm{1+\int_0^1 \di z \; \ex{z\widehat A}
 \widehat B \, \ex{-z\widehat A}}\ex{\widehat A},
\ee
where the subscript LE stands for Local Equilibrium and implies the calculation of 
mean values with the local equilibrium density operator (see Sect.~\ref{mean}). 
Thereby, putting together (\ref{expa1}) and (\ref{expa2}) and retaining only 
first-order terms in $\widehat B$:
$$
 \dens \simeq \brackets{1 - \aveql{\widehat B}}\denseql + 
 \int_0^1\di z \; \ex{z\widehat A}\widehat B \ex{-z\widehat A}\denseql,
$$
hence the mean value of an operator $\obs(y)$ becomes:
\be\label{meanobs}
 \ave{\obs(y)} \simeq \brackets{1-\aveql{\widehat B}}\aveql{\obs(y)} + 
 \Big\langle \obs(y)\int_0^1 \di z \; \ex{z\widehat A}\widehat B \ex{-z\widehat A}
 \Big\rangle.
\ee
Let us focus on the last term, which, by virtue of (\ref{bterms}), contains
expressions of this sort:
$$
 \aveql{\obs(y) \wX'(z,t,{\bf x})} \equiv \aveql{\obs(y) \, \ex{z\widehat A}
 \wX(t,{\bf x}) \, \ex{-z\widehat A}}
$$
where $\wX$ stands for components of either $\wT$ or $\wspt$ or $\partial_0\wspt$. 
From the identity:
$$
 \aveql{\obs(y)\wX'(z,t,{\bf x})} = \int_{-\infty}^{t} \diff\tau \;
 \aveql{\obs(y)\partial_\tau \wX'(z,\tau,{\bf x})} + 
 \lim_{\tau\to-\infty} \aveql{\obs(y)\wX'(z,\tau,{\bf x})},
$$
and the observation that correlations vanish for very distant times (check footnote
\ref{lcone}), one obtains:
\be\label{bcontent}
 \aveql{\obs(y)\wX'(z,t,{\bf x})} = \int_{-\infty}^{t} \diff\tau \;
 \aveql{\obs(y)\partial_\tau \wX'(z,\tau,{\bf x})} +
 \lim_{\tau\to-\infty} \aveql{\obs(y)}\aveql{\wX(\tau,{\bf x})},
\ee
where we have also taken advantage of the commutation between $\exp[\widehat A]$
and $\exp[\pm z \widehat A]$.
 
We now approximate \cite{hosoya} the local equilibrium density operator with
the nearest equilibrium operator $\wrho_0$ in Eq.~(\ref{rhoeq}), which also implies
that:
$$
  \widehat A \simeq -\widehat H/T 
$$
where $\widehat H$ is the hamiltonian operator (which ought to exists given the 
chosen boundary conditions). The straightforward consequence of this approximation 
is that the second term on the right hand side in Eq.~(\ref{bcontent}) can be written as:
$$
\aveql{\wX(-\infty,{\bf x})} \simeq \aveq{\wX(-\infty,{\bf x})} = 
\aveq{\wX(t,{\bf x})} 
$$
because the mean value is stationary under the equilibrium distribution. Therefore,
the Eq.~(\ref{bcontent}) can be approximated as:
\be\label{bcontent2}
 \aveql{\obs(y)\wX'(z,t,{\bf x})} \simeq \int_{-\infty}^{t} \diff\tau \;
 \aveq{\obs(y)\partial_\tau \wX'(z,\tau,{\bf x})} + \aveq{\obs(y)}
 \aveq{\wX(t,{\bf x})},
\ee
and the (\ref{meanobs}) as:
\be\label{meanobs2}
 \ave{\obs(y)} \simeq (1-\aveq{\widehat B})\aveq{\obs(y)} + 
 \int_0^1 \di z \; \aveq{\obs(y) \, \ex{-z\widehat H/T}\widehat B \, 
 \ex{z\widehat H/T}} 
\ee
Once integrated, the second term in (\ref{bcontent2}) gives rise to a term which
cancels out exactly the $\aveq{\widehat B}\aveq{\obs(y)}$ in the equation above,
which then becomes:
\be\label{meanobs3}
 \ave{\obs(y)} \simeq \aveq{\obs(y)} + \int_0^1 \di z \int_{-\infty}^{t} \diff\tau 
 \; \aveq{\obs(y)\partial_\tau \wX'(z,\tau,{\bf x})}
\ee
Let us now integrate the last term on the right hand side in $z$:
$$
 \int_0^1 \di z \int_{-\infty}^{t} \diff\tau \; \aveq{\obs(y)\partial_\tau \wX'(z,\tau,{\bf x})}
 = \frac{1}{\betabar}\int_0^\betabar\di u \int_{-\infty}^{t} \di \tau \; 
 \aveq{\obs(y)\partial_\tau \e^{-u \widehat H} \wX(\tau,{\bf x}) \e^{u\widehat H} }
$$
where $\betabar=1/T$ and $\betabar z=u$. As $\widehat H$ is the generator of 
time translations:
\begin{eqnarray*}
 && \frac{1}{\betabar}\int_0^\betabar\di u \int_{-\infty}^{t} \di \tau \; 
 \aveq{\obs(y)\partial_\tau \e^{-u \widehat H} \wX(\tau,{\bf x}) \e^{u\widehat H}}
 = \frac{1}{\betabar}\int_0^\betabar\di u \int_{-\infty}^{t} \di \tau \; 
 \aveq{\obs(y)\partial_\tau \wX(\tau+\ii u,{\bf x})} \nonumber \\
 &=& \frac{1}{\ii \betabar}\int_0^\betabar\di u \int_{-\infty}^{t} \di \tau \; 
 \aveq{\obs(y)\frac{\partial}{\partial u} \wX(\tau+\ii u,{\bf x})}
 = \frac{1}{\ii \betabar}\int_0^\betabar\di u \int_{-\infty}^{t} \di \tau \; 
 \frac{\partial}{\partial u} \left( \aveq{\obs(y) \wX(\tau+\ii u,{\bf x})} \right) 
 \nonumber \\
 &=& \frac{1}{\ii \betabar}\int_{-\infty}^{t} \di \tau \int_0^\betabar\di u \;
 \frac{\partial}{\partial u} \left( \aveq{\obs(y) \wX(\tau+\ii u,{\bf x})} \right) 
 = \frac{1}{\ii \betabar}\int_{-\infty}^{t} \left( \aveq{\obs(y)\wX(\tau+\ii\betabar,
 {\bf x})} - \aveq{\obs(y)\wX(\tau,{\bf x})}\right)
\end{eqnarray*}
On the other hand:
\begin{eqnarray*}
 && \aveq{\obs(y)\wX(\tau+\ii\betabar,{\bf x})} = \tr (\wrho_0 
  \obs(y) \e^{-\betabar \widehat H} \wX(\tau,{\bf x}) \e^{+\betabar \widehat H})
  = \frac{1}{Z_0} \tr(\e^{-\betabar \widehat H} \obs(y) \e^{-\betabar \widehat H} 
  \wX(\tau,{\bf x}) \e^{\betabar \widehat H}) \nonumber \\
 && = \frac{1}{Z_0} \tr(\obs(y) \e^{-\betabar \widehat H} 
 \wX(\tau,{\bf x})) = \tr(\wX(\tau,{\bf x})) \wrho_0 \obs(y)) = 
 \aveq{\wX(\tau,{\bf x})\obs(y)} 
\end{eqnarray*}
Hence, putting the last three equations together, we have:
\be
\int_0^1 \di z \int_{-\infty}^{t} \diff\tau \; \aveq{\obs(y)\partial_\tau 
\wX'(z,\tau,{\bf x})} = \frac{1}{\ii \betabar}\int_{-\infty}^{t} \di \tau \;
 \aveq{[\wX(\tau,{\bf x}),\obs(y)]} 
\ee
Substituting now $\wX$ with its specific operators, Eq.~(\ref{meanobs3}) can be
expanded as:
\bea\label{meanobs4}
 && \delta \ave{\obs(y)} = \ave{\obs(y)} - \aveq{\obs(y)} \simeq 
 \lim_{\varepsilon\to 0} \; \frac{1}{\ii\betabar} \int_{-\infty}^{t'} \di t \;
 \ex{\varepsilon(t-t')} \int_{-\infty}^t \di \tau \int \di^3 \x \;  
 \aveq{\comm{\wT^{\mu\nu}(\tau,{\bf x}),\obs(y)}}\partial_\mu\deltabeta_\nu(x) \nonumber \\
 && - \frac{1}{2} \lim_{\varepsilon\to 0} \frac{1}{i\betabar} \int_{-\infty}^{t'} \di t \;
 \ex{\varepsilon(t-t')} \frac{\partial}{\partial t} \int_{-\infty}^{t} \di \tau 
 \int \di^3 \x \; \aveq{\comm{\wspt^{0,\mu\nu}(\tau,{\bf x}),\obs(y)}} \delta\omega_{\mu\nu}(x)
 \nonumber \\
 && = \lim_{\varepsilon\to 0} \; \frac{1}{\ii\betabar} \int_{-\infty}^{t'} \di t \;
 \ex{\varepsilon(t-t')} \int_{-\infty}^t \di \tau \int \di^3 \x \;  
 \aveq{\comm{\wT^{\mu\nu}(\tau,{\bf x}),\obs(y)}}\partial_\mu\deltabeta_\nu(x) \nonumber \\
 && - \frac{1}{2} \lim_{\varepsilon\to 0} \frac{1}{i\betabar} \int_{-\infty}^{t'} \di t \;
 \ex{\varepsilon(t-t')} \int \di^3 \x \; 
 \aveq{\comm{\wspt^{0,\mu\nu}(t,{\bf x}),\obs(y)}} \delta\omega_{\mu\nu}(x) \nonumber \\
 && - \frac{1}{2} \lim_{\varepsilon\to 0} \frac{1}{i\betabar} \int_{-\infty}^{t'} \di t \;
 \ex{\varepsilon(t-t')}  \int_{-\infty}^{t} \di \tau \int \di^3 \x \; 
 \aveq{\comm{\wspt^{0,\mu\nu}(\tau,{\bf x}),\obs(y)}} \frac{\partial}{\partial t}
 \delta\omega_{\mu\nu}(x)
\end{eqnarray}
The first term on the right hand side of the above equation can be integrated by 
parts using:
\begin{eqnarray*}
&& \int_{-\infty}^{t'} \di t \; \ex{\varepsilon(t-t')} \int_{-\infty}^{t} \di \tau \; f(\tau) 
= \int_{-\infty}^{t'} \di t \; \frac{\partial}{\partial t} 
\brackets{\frac{\ex{\varepsilon(t-t')}}{\varepsilon}} \int_{-\infty}^{t}\di \tau \; 
f (\tau) \nonumber \\
&& = \frac{1}{\varepsilon}\int_{-\infty}^{t'} \di \tau \; f(\tau) - 
\int_{-\infty}^{t'} \di t \; \frac{\ex{\varepsilon(t-t')}}{\varepsilon} f(t) 
= \int_{-\infty}^{t'} \di t \frac{1-\ex{\varepsilon(t-t')}}{\varepsilon} f(t)
\end{eqnarray*}
so that the Eq.~(\ref{meanobs4}) can be finally written:
\bea\label{meanobsf}
 \delta \ave{\obs(y)} &=& 
 \lim_{\varepsilon\to 0} \; \frac{1}{\ii\betabar} \int_{-\infty}^{t'} \di t \;
 \frac{1-\ex{\varepsilon(t-t')}}{\varepsilon} \int \di^3 \x \;  
 \aveq{\comm{\wT^{\mu\nu}(x),\obs(y)}}\partial_\mu\deltabeta_\nu(x) \nonumber \\
 && - \frac{1}{2} \lim_{\varepsilon\to 0} \frac{1}{i\betabar} \int_{-\infty}^{t'} \di t \;
 \ex{\varepsilon(t-t')} \int \di^3 \x \; 
 \aveq{\comm{\wspt^{0,\mu\nu}(x),\obs(y)}} \delta\omega_{\mu\nu}(x) \nonumber \\
 && - \frac{1}{2} \lim_{\varepsilon\to 0} \frac{1}{i\betabar} \int_{-\infty}^{t'} \di t \;
 \ex{\varepsilon(t-t')}  \int_{-\infty}^{t} \di \tau \int \di^3 \x \; 
 \aveq{\comm{\wspt^{0,\mu\nu}(\tau,{\bf x}),\obs(y)}} \frac{\partial}{\partial t}
 \delta\omega_{\mu\nu}(x)
\eea

Another useful (equivalent) expression of $\delta \ave{\obs(y)}$ can be obtained
starting from the expression (\ref{elabor2}) of $\wUps$, where the continuity equation
for angular momentum is used from the beginning. Repeating the same reasoning
as above, it can be shown that one gets:
\bea\label{meanobsf2}
 \delta \ave{\obs(y)} &=& 
 \lim_{\varepsilon\to 0} \; \frac{1}{2\ii\betabar} \int_{-\infty}^{t'} \di t \;
 \frac{1-\ex{\varepsilon(t-t')}}{\varepsilon} \int \di^3 \x \;  
 \aveq{\comm{\wT^{\mu\nu}_S(x),\obs(y)}}(\partial_\mu\deltabeta_\nu(x) + 
 \partial_\nu\deltabeta_\mu(x))\nonumber \\
 && + \lim_{\varepsilon\to 0} \frac{1}{2\ii \betabar} \int_{-\infty}^{t'} \di t \;
 \frac{1-\ex{\varepsilon(t-t')}}{\varepsilon} \int \di^3 \x \; 
 \aveq{\comm{\wT^{\mu\nu}_A(x),\obs(y)}} (\partial_\mu\deltabeta_\nu(x) - 
 \partial_\nu\deltabeta_\mu(x) + 2\delta\omega_{\mu\nu}(x)) \nonumber \\
 && - \frac{1}{2} \lim_{\varepsilon\to 0} \frac{1}{i\betabar} \int_{-\infty}^{t'} \di t \;
 \ex{\varepsilon(t-t')}  \int_{-\infty}^{t} \di \tau \int \di^3 \x \; 
 \aveq{\comm{\wspt^{\lambda,\mu\nu}(\tau,{\bf x}),\obs(y)}} \partial_\lambda\delta
 \omega_{\mu\nu}(x)
\eea
As we have pointed out, these expressions hold when $\wrho_0$ has a non-vanishing 
chemical potential.

\section*{APPENDIX B - Commutators and discrete symmetries}

We want to study the effect of space inversion and time reversal on the mean value
of commutators like:
$$
 \aveq{\comm{\obs_1^{\mu_1 \cdots \mu_m}(t,\x), \obs_2^{\nu_1\cdots\nu_n}(0,{\bf 0})}}
$$
where $\obs_1$ and $\obs_2$ are physical tensor densities of rank $m$ and $n$,
respectively.

The equilibrium density operator $\wrho = \exp[-\widehat H/T]/Z$ is symmetric for
space-time translations and rotations, as well as time reversal and parity if the 
hamiltonian is itself parity and time reversal invariant. The symmetry under this 
class of transformations allows to simplify the above expression.
For any linear unitary transformation $\widehat {\sf U}$ which commutes with $\wrho$ 
one has:
$$
 \aveq{\obs} = \trace{\dens_0 \, \obs} = \trace{\widehat {\sf U}^{-1}\dens_0\widehat 
 {\sf U}\; \obs}  = \trace{\dens_0\,\widehat {\sf U}\obs\widehat {\sf U}^{-1}} = 
 \aveq{ \widehat {\sf U}\obs\widehat {\sf U}^{-1} }
$$
Taking $\widehat {\sf U}=\widehat {\sf T}(a)$ with $\widehat{\sf T}(a)$ a general
translation operator:
$$
 \aveq{\comm{\obs_1^{\mu_1 \cdots \mu_n}(t,{\bf x}),\obs_2^{\nu_1\cdots\nu_n}
 (0,{\bf 0})}} = \aveq{\comm{\obs_1^{\mu_1 \cdots \mu_n}(t + a^0,{\bf x}+{\bf a}), 
 \obs_2^{\nu_1\cdots\nu_n}(a^0,{\bf a})}}
$$
and so, setting $(a^0,{\bf a})=(-t,-{\bf x})$:
$$
 \aveq{\comm{\obs_1^{\mu_1 \cdots \mu_m}(t,{\bf x}),\obs_2^{\nu_1\cdots\nu_n}(0,{\bf 0})}} 
 = \aveq{\comm{\obs_1^{\mu_1 \cdots \mu_m}(0,{\bf 0}), \obs_2^{\nu_1\cdots\nu_n}(-t,-{\bf x})}}
$$
Similarly, for a space inversion:
$$
  \aveq{\comm{\obs_1^{\mu_1 \cdots \mu_m}(t,{\bf x}),\obs_2^{\nu_1\cdots\nu_n}(0,{\bf 0})}} 
 = (-1)^{n_s+m_s}\aveq{\comm{\obs_1^{\mu_1 \cdots \mu_m}(t,-{\bf x}),
  \obs_2^{\nu_1\cdots\nu_n}(0,{\bf 0})}} 
$$
where $m_s$ and $n_s$ are the number of space indices among $\mu_1,\cdots\mu_m$ and 
$\nu_1,\cdots \nu_n$ respectively. 

The time reversal operator $\widehat\Theta$ is antiunitary, thus a point-dependent 
physical scalar operator $\widehat A(t,{\bf x})$ transforms as follows:
$$
  {\widehat \Theta} \widehat A(t,{\bf x}) {\widehat \Theta}^{-1}
  = \widehat A^\dagger(-t,{\bf x})
$$
whence, for commutators:
$$
 {\widehat \Theta}\comm{\widehat A(t,{\bf x}),\widehat B(t,{\bf x})}{\widehat \Theta}^{-1}=
 \comm{\widehat B^\dagger(-t,{\bf x}),\widehat A^\dagger(-t,{\bf x})}
$$
Then, for Hermitian operators, what gets changed is the order of the operators besides 
their time argument. For tensor hermitian observables and time-reversal symmetric 
hamiltonian, one obtains:  
$$
  \aveq{\comm{\obs_1^{\mu_1 \cdots \mu_m}(t,{\bf x}), 
  \obs_2^{\nu_1\cdots\nu_n}(0,{\bf 0})}} =  
  (-1)^{m_0+n_0}\aveq{\comm{ \obs_2^{\nu_1\cdots\nu_n}(0,{\bf 0}), 
  \obs_1^{\mu_1 \cdots \mu_m}(-t,{\bf x})}}
$$
where $m_0$ and $n_0$ are the number of time indices among $\mu_1,\cdots\mu_m$ and 
$\nu_1,\cdots \nu_n$ respectively.

\end{document}